\documentclass[11pt]{article}

\usepackage{subcaption,lipsum,graphics,tikz,authblk,fancyhdr,natbib,amsmath,amsthm,amsopn,amstext,amsfonts,amssymb,enumitem,multirow,standalone,url}
\usepackage[a4paper, margin=1in, headheight=13.6pt]{geometry}
\graphicspath{{./figures/}}
\usetikzlibrary{shapes, arrows,decorations,automata,backgrounds,petri,positioning,arrows}

\title{New Formulations and Pricing Mechanisms for Stochastic Electricity Market Clearing Problem}
\author[1]{Sakitha~Ariyarathne\thanks{sariyarathne@smu.edu}}
\author[1]{Harsha~Gangammanavar\thanks{harsha@smu.edu}}
\affil[1]{Department of Operations Research and Engineering Management, Southern Methodist University, Dallas TX}

\newenvironment{history}[1][History]{%
	\begin{quotation} \noindent\it #1:
}{%
  \end{quotation} 
}
\newtheorem{theorem}{Theorem}[section]
\newtheorem{lemma}[theorem]{Lemma}
\newtheorem{proposition}[theorem]{Proposition}

\newtheorem{assumption}[theorem]{Assumption}

\theoremstyle{definition}
\newtheorem{definition}{Definition}[section]

\theoremstyle{remark}

\newcommand{\rvset}{\Xi}
\newcommand{\rv}{\tilde{\xi}}
\newcommand{\obs}{\xi}
\newcommand{\set}[1]{\mathcal{#1}}
\newcommand{\RR}{\mathbb{R}}
\newcommand{\expect}[2]{\mathbb{E}_{#2}[#1]}
\newcommand{\Expect}[2]{\mathbb{E}_{#2}\bigg[#1\bigg]}
\newcommand{\inner}[1]{\langle #1 \rangle}
\def\argmin{\mathop{\rm arg\,min}}%
\def\argmax{\mathop{\rm arg\,max}}%

\newif\ifpaper
\paperfalse

\begin{document}\thispagestyle{empty}
\maketitle

\begin{abstract}
We present new formulations of the stochastic electricity market clearing problem based on the principles of stochastic programming. Previous analyses have established that the canonical stochastic programming model effectively captures the relationship between the day-ahead and real-time dispatch and prices. The resulting quantities exhibit desirable guarantees of revenue adequacy, cost recovery, and price distortion in expectation. Motivated by these results, we develop alternative stochastic programming model formulations that differ in how we model the nonanticipativity requirement. We design new pricing mechanisms by developing suitable dual optimization problems for these models. We analyze these pricing mechanisms to assess their ability to provide revenue adequacy for the system operators, cost recovery for generators, and price distortion. Unlike the previous mechanisms that yield the desired properties in expectation, one of our pricing mechanisms extends the benefits of cost recovery and bounded price distortion under every scenario while retaining the long-run revenue adequacy for the system operator. We demonstrate these benefits through numerical experiments conducted on well-known test systems.\end{abstract}
\begin{history}
First submission: May, 2023; Current version \today.
\end{history}

\renewcommand{\obs}{\omega}
\renewcommand{\rv}{\tilde{\omega}}
\renewcommand{\rvset}{\Omega}
\newcommand{\x}{\mathbf{x}}
\newcommand{\y}{\mathbf{y}}

\newcommand{\hg}[1]{\textcolor{purple}{HG: #1}}
\newcommand{\sa}[1]{\textcolor{blue}{SA: #1}}

\section{Introduction}
Power systems operators extensively use deterministic optimization to decide the electricity market's operational decisions, such as unit commitment and dispatch. These decisions are made using bids for supply and demand submitted by generating firms and utilities. In addition to clearing the market, the solutions obtained from deterministic optimization are also used to form electricity prices. A conventional arrangement involves day-ahead and real-time markets. This competitive market arrangement facilitates operational decisions and produces prices that enable efficient utilization of energy resources and incentivizes participation (see \cite{ott2003experience}).  

The typical operations decision process involves solving multiple layers of unit commitment problems (e.g., day-ahead, short-term, and real-time) and culminates by solving an economic dispatch (ED) problem. However, the pricing mechanisms used in practice and those proposed in academic literature often apply just the day-ahead and real-time components. A standard procedure involves solving the unit commitment problem first, followed by a market-clearing problem composing only the committed units. The optimal dual multiplier of the flow-balance equation, which ensures sufficient electricity supply is available to meet the total demand at each power network node, is used as the electricity price at the node. These dual multipliers are often called locational marginal prices (LMPs). 

This standard practice of price formation provides equilibrium prices that support generation costs for committed units in the deterministic convex case. However, there are two notable shortcomings. Firstly, this practice fails when the optimization problem involves nonconvexity, for instance, from introducing binary unit commitment variables. Consequently, the standard price mechanisms do not support fixed components of the generator's operating costs, such as start-up and no-load costs. Secondly, they fail to explicitly capture the effect of uncertainty on price formation. This paper attempts to advance the research concerning the latter shortcoming of the standard practice. 

Stochastic programming (SP) provides a systematic framework to study decision-making under uncertainty \citep{birge2011introduction}. SP models and methods have effectively addressed many power systems planning and operations problems. The advent of large-scale integration of renewable resources for electricity generation has further enhanced the interest in applying SP-based approaches. The solutions obtained from these approaches have economic and environmental benefits while improving operational reliability \citep{atakan2022towards} While there is an established recognition for using SP for power systems operations problems, tools offered by SP have yet to be fully exploited for electricity price formation. 	

Irrespective of the approach adopted for the price mechanism, the resulting prices must satisfy properties that encourage market participation; these include revenue adequacy for the system operator and cost recovery for generators. The former ensures the system operator generates sufficient revenue to pay the generators, avoiding a deficit. The latter ensures that the generators recover their costs through the payments they receive. Notice that if a generator fails to achieve cost recovery, side or uplift payments become necessary to incentivize participation in the market.

\subsection{Previous work on SP-based Price Formation}
The effect of uncertainty in spot (real-time) pricing was first studied using a simulation-based approach by \cite{kaye1990forward}. This work establishes forward (day-ahead) contracts to share the risk between the demand and supply sides. The authors propose to set the day-ahead price to the best estimate (expectation) of the real-time price. \cite{wong2007pricing} were among the first to offer pricing mechanisms based on solutions of two-stage SP models. In their model, the first stage is a day-ahead capacity market, and the second stage includes energy and reserve utilization in the real-time market. The prices are based only on second-stage flow-balance equations. The authors propose using the expected value of the real-time prices as the day-ahead price. However, this approach fails to achieve the desired price-formation properties. To overcome this deficiency, the authors propose certain adjustments to their price formation without establishing if these adjustments support market equilibrium. 

\cite{pritchard2010single} use a two-stage SP model that gives prices that reflect uncertainty. The first stage in their model involves day-ahead dispatch quantities and LMPs. The second stage captures the real-time corrections to these quantities and an LMP that reflects these corrections under every uncertain scenario. They demonstrate that such a pricing mechanism achieves revenue adequacy in expectation. Using a similar two-stage model and pricing mechanism, \cite{zavala2017stochastic} establish that generators recover their cost in expectation, or equivalently, the expected uplift payments are zero. They also introduce the notion of price distortion and argue that day-ahead prices need not be equal to the expected values of the real-time prices. They establish that the real-time adjustment bids (premiums) bound the difference between the day-ahead and expected real-time prices. 

\cite{morales2012pricing} also use a two-stage linear SP formulation for the ED problem. Instead of real-time market clearing, they use the reserve to clear the deviation between day-ahead and real-time. The ramping capabilities of each generator determine reserve capacities. Using the pricing mechanism of \cite{pritchard2010single}, the payment to the day-ahead capacity is made using the dual variable corresponding to the flow balance equation in the formulation, which does not include reserves, load shedding, and renewables. On the other hand, the payment to scenario-specific reserve utilization is based on reserve balancing (real-time flow balance) constraints. As before, the pricing mechanism achieves cost recovery and revenue adequacy in expectation since their first-stage decisions are scenario-independent.

\cite{zakeri2019pricing} propose that financial settlements be made only using second-stage real-time quantities and prices, while the first-stage solutions only serve operational needs. In a setting without a day-ahead market, the resulting pricing mechanism achieves revenue adequacy under each scenario while reaching cost recovery in expectation. Using a similar setting, \cite{cory2018payment} present an alternative pricing mechanism that achieves scenario-wise cost recovery while guaranteeing expected revenue adequacy. In all the pricing mechanisms presented so far, it is worthwhile to note that either cost recovery or revenue adequacy can be achieved under every scenario. At the same time, the other is ensured only in expectation. Simultaneously guaranteeing scenario-specific cost recovery and revenue adequacy comes at the expense of economic dispatch, as seen in \cite{kazempour2018stochastic}. The undesirable cost of market inefficiency is imposed on the loads, as is the practice.

\subsection{Contributions}
The setting of \cite{pritchard2010single} and the corresponding pricing mechanism achieves both revenue adequacy and cost recovery in expectation while ensuring an optimal dispatch. Previous studies ensure cost recovery or revenue adequacy under every scenario while achieving the other in expectation. Furthermore, the value of a two-stage decision process is well established in its ability to identify operational decisions that are well-hedged against uncertainty and maintain computational viability. Therefore, in this paper, we revisit the two-stage setting of \cite{pritchard2010single} to develop our pricing mechanisms. In light of the above observations regarding the existing stochastic clearing problem models and the pricing mechanisms derived from them, the main contributions of this study are as follows.
\begin{enumerate}
    \item \emph{New pricing mechanisms.} We develop two alternative models for the two-settlement electricity market that account for renewable generation and demand uncertainty. These models are based on two-stage SP principles, particularly on the notion of nonanticipativity. We develop two models based on the alternative approaches to capture the nonanticipativity requirement: mean-vector and state-vector formulations. In addition, we introduce two pricing mechanisms using solutions of these models and show the relationships between the models using stochastic convex programming duality. %We show that both pricing mechanisms have improved pricing characteristics over the pricing mechanism proposed in \cite{pritchard2010single}.
	\item \emph{Analyses of pricing mechanisms.} We show that the new pricing mechanisms achieve revenue adequacy for the system operator in expectation; that is, the system operator does not face a financial deficit in the long run. We also prove that every participating generator achieves cost recovery without any out-of-market payments in expectation with mean-vector formulations' pricing mechanism. For the state-vector formulations' pricing mechanism, we prove that all generators achieve cost recovery under every scenario, a more significant result. Finally, for the state-vector formulation-based pricing mechanism, we develop bounds on scenario-specific price distortion, the deviation between first- and second-stage unit prices under every scenario. We illustrate these properties through numerical experiments.
\end{enumerate}

The paper is structured as follows. In the following section \S\ref{sec:marketSetting}, we describe the market setting we consider in this study, introduce the stochastic market clearing model elements, and the market metrics.  We present the alternative SP models for the stochastic market clearing problem in \S\ref{sec:pricing_mechanisms_analysis}. We also present the pricing mechanisms and their analyses in this section. The proofs for all the results are presented in the associated online supplement. Finally, in section \S\ref{sec:expriments}, we numerically illustrate the properties of different pricing mechanisms introduced in the previous sections.

\section{Market Setting} \label{sec:marketSetting}
In this section, we present the electricity market-clearing model formulations. We begin by describing the market setting we consider in our study. We also present the metrics used to analyze and compare the pricing mechanisms.

We consider a two-settlement market setting where the market participants partake in a forward day-ahead market and a real-time balancing market. An Independent System Operator (ISO) manages and clears the markets centrally. By $\set{P}$, we denote the market participants who make offers separately in the day-ahead and real-time markets. We assume that a market participant operates a single unit and does not cooperate with other market participants, thus ensuring perfect competition. Therefore, a participant either operates a generation unit or is a demand aggregator. We denote by $\set{G}$ and $\set{D}$ the set of generators and loads, respectively, with $\set{P} = \set{G} \cup \set{D}$. Typically, the ISO clears the day-ahead market by first scheduling generators using a security-constrained unit commitment problem. Following the commitment problem, the ISO identifies optimal generation amounts and reserves using a security-constrained economic dispatch problem (see, for example, \cite{pjmManual2022} for details regarding market operations at PJM ISO). Our formulations assume that the unit commitment decisions are made a priori and are available as input. This market setting is also the subject of previous works on stochastic market clearing, particularly in \cite{pritchard2010single} and \cite{zavala2017stochastic}.

\subsection{Model Elements} \label{subsec:modelElements}
The power network is captured by a set of nodes (buses) $\set{N}$ and a set of lines $\set{L}$. The day-ahead market is cleared using the day-ahead offers that include price bids $c_i$ for $i \in \set{P}$. The offers also include estimates of available maximum/minimum capacity that we denote by $x^{\max}_i/x^{\min}_i$ for $i \in \set{P}$. We denote the day-ahead settlement/cleared amounts by the decision variable $x_i$ for either type of participants $i \in \set{P}$. A decision variable corresponding to a generator satisfies $x_i \geq 0$, and that corresponding to a demand satisfies $x_i \leq 0$. These constitute the principal elements of the consolidated first-stage decision vector $\y_1 \in \RR^{n_1}$. In addition to the day-ahead clearing decisions, the decision vector $\y_1$ also includes the power flows $f_{ij}$ for $(i,j) \in \set{L}$ that support the cleared day-ahead quantities. Therefore, we define $\y_1 := ((x_i)_{i \in \set{P}}, (f_{ij})_{(i,j) \in \set{L}})$. The capacity limits on these decision variables define the set $\set{C}_1$ as:
\begin{align}
\label{eq:da_detSet}
    \set{C}_1 := \left \{ \y_1 ~\left \vert 
    	\begin{array}{l}
    		x^{\min}_i \leq x_i \leq x^{\max}_i \qquad \forall i \in \set{P}, \\
    		f^{\min}_{ij} \leq f_{ij} \leq f^{\max}_{ij} \qquad \forall (i,j) \in \set{L}, \\
    		(f_{ij})_{(i,j) \in \set{L}} \in \set{F}.
    	\end{array}\right. \right \}
\end{align}
The first two constraints limit the first-stage decisions within their capacity limits. These limits capture a generator's minimum up-time generation (lower bound ) and generation capacity (upper bound). On the other hand, for a load, the lower bound is set to the negative of the demand forecast, and the upper bound is set to zero. The third requirement ensures that the power flows satisfy the underlying physics. The description of the set $\set{F}$ depends on the nature of approximation (such as the direct-current approximation) or relaxations (such as second-order conic programming and semidefinite programming-based relaxations) adopted to describe the power flows on a line. We refer the reader to \cite{molzahn2019survey} for a detailed review on modeling power flows. While choosing a particular form of $\set{F}$ has computational implications, our discussion in this paper is independent of the choice. Irrespective of the representation chosen, we assume that $0 \in \set{F}$. In addition to the above, the cleared day-ahead quantities and the power flows satisfy the flow balance at all nodes. These are given by
\begin{align}
\label{eq:da_flowBalance_original}
    \underbrace{\sum_{j:(j,i) \in \set{L}}f_{ji} - \sum_{j:(i,j) \in \set{L}}f_{ji}}_{:= \tau_i(f)} + \sum_{j \in \set{P}(i)}x_j = 0 \qquad \forall i \in \set{N}.
\end{align}
The function $\tau_i(f) = \sum_{j:(j,i) \in \set{L}}f_{ji} - \sum_{j:(i,j) \in \set{L}}f_{ji}$ captures the net flow (difference between inflow and outflow) at bus $i$. For ease of exposition, we adopt an abstract form for power flow equations as $g_1(\y_1) = 0$. Finally, the day-ahead social deficit (negative social surplus, formally defined later in this section ) is given by
\begin{align} \label{eq:da_obj}
    f_1(\y_1) := \sum_{i \in \set{P}} c_ix_i.
\end{align}
This constitutes the first-stage objective function.

The day-ahead and the real-time markets are cleared sequentially. The real-time market conditions are uncertain when the day-ahead market is cleared. The uncertain conditions include (but are not limited to) renewable generation, demand, transmission line capacities, and generator status. These uncertain elements are modeled by the random vector $\rv$ defined on a probability space $(\Omega, \set{F}, \mathbb{P}),$ where $\Omega$ is a set of all possible scenarios, $\set{F}$ is the sigma-algebra, and $\mathbb{P}$ is the probability distribution. We do not restrict $\Omega$ to be a finite set. The real-time market is cleared only after observing a realization $\obs$ of the random vector $\rv$. We will refer to individual realizations as scenarios. Therefore, the quantities cleared in the real-time market may deviate from those cleared in the day-ahead market. 

The market participants submit separate generation and demand bids to address the deviation between the day-ahead and real-time conditions. The real-time offers include the additional cost (premiums) for positive $(\delta^{+}_i)_{i \in \set{P}}$ and negative deviations $(\delta^{-}_i)_{i \in \set{P}}$ incurred by the market participants. They also include maximum and minimum capacity bounds given by $X^{\max}_i/X^{\min}_i$ for $i \in \set{P}$. Analogous to the day-ahead settlements, we denote real-time settlements/cleared amounts under realization $\obs$ by $X_i(\obs)$ for $i \in \set{P}$ that satisfy $X_i(\obs) \geq 0$ for generators and $X_i(\obs) \leq 0$ for loads. We collectively denote the real-time decisions by the vector $\y_2(\obs) := ((X_i(\obs))_{i \in \set{P}}, (F_{ij}(\obs))_{(i,j) \in \set{L}}) \in \RR^{n_2}$, where $F_{ij}(\obs)$ for $(i,j) \in \set{L}$ denote the power flows that support real-time amounts. In the real-time market, the day-ahead amount of a participant $i \in \set{P}$ is adjusted by either $(X_i(\obs) - x_i)_+$ or $(X_i(\obs) - x_i)_-$, incurring operations cost $\delta_i^{+}$ and $\delta_i^{-}$, respectively. These deviation costs are a small fraction of the day-ahead cost $c_i$. Therefore, the day-ahead cost is adjusted by 
\begin{align} \label{eq:rt_obj}
    f_2(\y_1,\y_2(\obs),\obs) := &\sum_{i \in \set{P}} \bigg[ (c_i+ \delta^{+}_i)(X_i(\obs)-x_i)_+ - (c_i-\delta^{-}_i)(X_i(\obs) - x_i)_- \bigg].
\end{align}
The above constitutes the second-stage objective function. We define the real-time counterparts of $\set{C}_1$ and flow-balance equations for $\obs \in \Omega$ as
\begin{align}
\label{eq:rt_detSet}
    \set{C}_2(\obs) := \left \{ \y_2(\obs) \left \vert 
    	\begin{array}{l}
    		\max\{X^{\min}_i,-X^{\text{avail}}_i(\obs)\} \leq X_i(\obs) \leq \min\{X^{\max}_i, X^{\text{avail}}_i(\obs)\} \qquad \forall i \in \set{P},\\
    		F^{\min}_{ij} \leq F_{ij}(\obs) \leq F^{\max}_{ij} \qquad \forall (i,j) \in \set{L},\\
            (F_{ij}(\obs))_{(i,j) \in \set{L}} \in \set{F}
    	\end{array}\right. \right \}
\end{align}
and 
\begin{align}
\label{eq:rt_flowBalance}
    \tau_i(F(\obs)) - \tau_i(f(\obs)) + \sum_{j \in \set{P}(i)} X_j(\obs) - x_j(\obs) = 0 \qquad \forall i \in \set{N},
\end{align}
respectively. As before, we will use  $g_2(\y_2(\obs)) = 0$ to represent the flow-balance equations succinctly. The quantity $X_i^{\text{avail}}(\obs)$ used in \eqref{eq:rt_detSet} denotes the observed quantity of $i^{th}$ participant under realization $\obs$. Notice the explicit dependence of the cleared real-time quantities, power flows, and the cost function upon the realization $\obs$.

\subsection{Market Metrics} \label{subsec:marketMetrics}
The next section presents pricing mechanisms based on the alternative SP formulations and their properties. For each pricing mechanism, we analyze the payment received by market participants and the revenue earned by the ISO. In our analyses, we use the notion of revenue adequacy for the ISO and cost recovery for the participating generators. Price consistency was identified as an appropriate metric for stochastic market settings in \cite{zavala2017stochastic} as they help achieve appropriate incentives. Therefore, we adopt price consistency as an additional metric to assess our pricing mechanisms. Before presenting the model formulations and pricing mechanisms, we introduce these metrics of interest.

Under a scenario $\obs$, a market participant realizes a value given by  
\begin{align} \label{eq:socialSurplus_part}
    \varphi_i(\obs) := - c_ix_i(\obs) - (c_i + \delta^+_i)(X(\obs) - x_i)_+ + (c_i - \delta^-_i)(X_i(\obs) - x_i)_- \qquad \forall i \in \set{P}.
\end{align}
Recall that $x_i, X_i \geq 0$ for generators, $x_i, X_i \leq 0$ for loads, and $\delta_i^+, \delta_i^- > 0$. Therefore, the realized value is negative for the generators and must be viewed as the generator's cost. On the other hand, the realized value is positive for loads which can be interpreted as the value gained by meeting their demand. The \emph{social surplus} is defined as the value realized across all the market participants under $\obs \in \Omega$. That is, 
\begin{align} \label{eq:socialSurplus}
    \varphi(\obs) := \sum_{i \in \set{P}}\varphi_i(\obs). 
\end{align}
The stochastic market clearing problem aims to minimize the negative of expected social surplus. The above form of social surplus can be derived from the definition of the day-ahead cost $f_1$ in \eqref{eq:da_obj}, and the real-time adjustment cost $f_2$ in \eqref{eq:rt_obj} through simple algebraic operations. It is worth noting that the expected social surplus is a systemwide measure viewed from the ISO's perspective. For individual participants, we denote by $\rho_i(\obs)$ the payment made to a generator (when $\rho_i(\obs) \leq 0$) by the ISO or received from a load (when $\rho_i(\obs) \geq 0$) under scenario $\obs$. 
\begin{definition} \label{def:cost_recovery}
    A participating generator $i$ achieves \emph{scenario-specific cost recovery} when,
    \begin{subequations}
    \begin{align}
        \rho_i(\obs) - \varphi_i(\obs) \leq 0,
    \end{align}
    under scenario $\obs$. The participating generator achieves \emph{cost recovery in expectation} when,
    \begin{align}
        \expect{\rho_i(\rv)}{} - \expect{\varphi_i(\rv)}{} \leq 0.
    \end{align}
    \end{subequations}
\end{definition}
Notice that the cost recovery metric is defined for only the participating generators. When a generator achieves cost recovery, it can cover its short-run costs through payments received from the ISO. This case, also known as making whole, encourages the generators to participate in the market. Without this feature, the payment mechanism has to incorporate additional side payments known as \emph{uplift payments}. Uplifts become essential under three cases (i) when approximations/relaxations are employed for the optimal power flow equations, (ii) the inclusion of binary-valued commitment decision, and (iii)  when an approximate representation of uncertainty is utilized. In this study, we focus our attention on the latter. From the ISO's perspective, the following metric is useful.
\begin{definition} \label{def:revenue_adequacy}
    The ISO achieves \emph{scenario-specific revenue adequacy} when,
    \begin{subequations}
    \begin{align}
        \sum_{i \in\set{P}}\rho_i(\obs) \geq 0,
    \end{align}
    under scenario $\obs$. The ISO achieves \emph{revenue adequacy in expectation} when,
    \begin{align}
        \Expect{\sum_{i \in\set{P}}\rho_i(\rv)}{} \geq 0.
    \end{align}
    \end{subequations}
\end{definition}
When we achieve revenue adequacy under a specific payment mechanism, the ISO receives sufficient payment from the loads to make payments to generators. It does not run into a financial deficit. Finally, we use the following definition to capture the deviation of price signals in the day ahead and in real-time.
\begin{definition}
    The \emph{scenario-specific price distortion}, denoted by $\mathbb{M}_n(\obs)$ for all $n \in \set{N}$ and $\obs \in \rvset$, is the difference between the day-ahead and real-time prices. Furthermore, the prices are said to be \emph{consistent in expectation} at a node $n$ if $\expect{\mathbb{M}_n(\rv)}{} = 0$.
\end{definition}
Notice that we introduce metrics in scenario-specific and expectation forms. While the scenario-specific statements are relatively stronger metrics, providing such strong guarantees may not always be possible, as we illustrate in the next section.

\section{Alternative Models and Pricing Mechanisms} \label{sec:pricing_mechanisms_analysis}
This section presents alternative SP problem formulations of the stochastic electricity market clearing problem. We present each formulation's primal form and develop a suitable dual. In identifying the alternative formulations, we particularly emphasize our computational ability to solve the formulation optimally. Before we present the SP models, we describe a deterministic model that will aid our discussions on pricing mechanisms in \S\ref{sec:pricing_mechanisms_analysis}. The models we present in this section satisfy the following assumption.
\begin{assumption} \label{assum:SCP}
The sets $\set{C}_1$ and $\set{C}_2$ are convex, closed, and nonempty. The functions $f_1(\cdot)$ on $\RR^{n_1}$ and $f_2(\cdot, \cdot, \obs)$ on $\RR^{n_1} \times \RR^{n_2}$ are convex, everywhere-defined, and finite. The latter function is measurable for each $(\y_1, \y_2) \in \RR^{n_1} \times \RR^{n_2}$. 
\end{assumption}
When we use the direct-current approximation or convex relaxations of power flows to represent $\set{F}$, the market clearing model elements satisfy the above assumptions. The real-time quantities must only be within specified bounds; see \eqref{eq:rt_detSet}, thereby allowing generation and load shedding in the second stage. This ensures that $f_2$ is everywhere-defined and finite as required by the above assumption. It is also worthwhile to note that, under Assumption \ref{assum:SCP}, the function $f_2(\cdot, \cdot, \rv)$ is proper with closed, measurable, and convex epigraph outside a subset of $\rvset$ with $\mathbb{P}$-measure zero.

\subsection{The Clairvoyant Problem} \label{subsec:clairvoyanProblem}
The clairvoyant problem is instantiated with the full knowledge of a scenario. For a scenario $\obs$, the total cost includes the cost realized in the day ahead followed by the cost associated with real-time balancing, i.e., $f_1(\y_1) + f_2(\y_1,\y_2,\obs)$. The clairvoyant problem is stated as follows
\begin{align} \label{eq:scenarioProblem}
	\min~& f_1(\y_1) + f_2(\y_1, \y_2, \obs) \tag{$\mathbf{S}_\obs$}\\
	\text{subject to}~& \y_1 \in \set{C}_1,~ g_1(\y_1) = 0, \notag \\
	& \y_2 \in \set{C}_2(\obs),~ g_2(\y_1, \y_2, \obs) = 0. \notag
\end{align}
Naturally, the optimal solution to the above problem is a function of scenario $\obs$ that we denote as $(\y_1^\star(\obs), \y_2^\star(\obs))$.  The problem \ref{eq:scenarioProblem} is often referred to as the ``scenario problem'' in the SP literature. Notice that the point forecast-based deterministic problems solved by the system operators in today's practice fall under our definition of a clairvoyant problem. Furthermore, if we use the mean vector $\bar{\obs} = \expect{\rv}{}$, the resulting problem is the so-called mean-value problem.

\subsection{Canonical Stochastic Programming Model} \label{subsubsec:canonicalModel}
The two-stage SP program for the two-settlement problem is typically stated as follows:
\begin{align}\label{eq:scp}
	\min~& f_1(\y_1) + \expect{f_2(\y_1, \y_2(\rv), \rv)}{\mathbb{P}} \tag{\bf C}\\
	\text{subject to}~& \y_1 \in \set{C}_1,~ g_1(\y_1) = 0,  \notag \\
	& \y_2(\rv) \in \set{C}_2,~ g_2(\y_1, \y_2(\rv), \rv) = 0 \qquad a.s. 	\notag
\end{align}
Notice the explicit dependence of the second-stage decision vector on the random variable $\rv$. Further, the second-stage constraints are required to hold almost surely, denoted as $a.s.$ (that is, these constraints must hold except for $\rv$ in a subset of $\rvset$ with $\mathbb{P}$-measure zero). Under Assumption \ref{assum:SCP}, the objective function in \eqref{eq:scp} is well-defined, and $\y_2$ is a bounded measurable function of $\rv \in \rvset$.

Let $H(\y_1,\obs) = \inf_{\y_2} \{f_2(\y_1, \y_2, \obs)~|~\y_2 \in \set{C}_2, g_2(\y_1, \y_2, \obs) = 0\}$. Using this, the intrinsic first-stage problem can be restated as:
\begin{align}\label{eq:scp_twoStage}
	\min~\{f_1(\y_1) + \expect{H(\y_1, \rv)}{\mathbb{P}} ~|~ \y_1 \in \set{C}_1, g_1(\y_1) = 0\}. \tag{\bf SP}
\end{align} 
Note that the objective function attains a value of $+\infty$ unless $f_1(\y_1) < +\infty$, \eqref{eq:scp} satisfies relatively complete recourse (i.e., $H(\y_1, \rv) < \infty$, almost surely), and $H(\y_1, \obs)$ is a measurable function of $\rv \in \rvset$ for each $\y_1 \in \RR^{n_1}$. If these conditions are met, a solution $\y_1$ attains minima to \eqref{eq:scp_twoStage} if and only if it attains minima to \eqref{eq:scp}. For results that establish this equivalence and the measurability of $h$, we refer the reader to Theorem 1 and Proposition 1 in \cite{rockafellar1976stochastic}, respectively. When $\rvset$ is finite, the above perspective on two-stage SP problems can be solved using the L-shaped method \citep{van1969shaped}. This decomposition-based approach provides a computationally viable path for large-scale implementation. The previous studies on stochastic market clearing problems (e.g., \cite{pritchard2010single} and \cite{zavala2017stochastic}) utilize this perspective. We refer the reader to \eqref{eq:canonical_model} in the online supplement \ref{subsec:appendix_canonicalForm} for a detailed presentation of the canonical model of the stochastic clearing problem.

\subsubsection{Pricing Mechanism 1}
The first pricing mechanism is based on the properties of the canonical form \eqref{eq:scp} of the SP problem when $\rvset$ is finite. This payment mechanism was proposed in \cite{pritchard2010single} and further analyzed in \cite{zavala2017stochastic}. We can interpret the dual solution $\pi_n^c$ corresponding to the first-stage flow-balance equation $(g_1(\y_1))_n = 0$ as the marginal cost of serving an additional unit of forecasted demand at node $n$. The second-stage dual solution $\Pi^c_n(\obs)$ can be interpreted as the marginal cost associated with adjustments in demand under scenario $\obs$. Therefore, the following payment mechanism is appropriate for the canonical form.
\begin{definition}
Under scenario $\obs \in \rvset$, the market participants receive a payment given by
\begin{align}\label{eq:Pricing_mechanism_C}
    \rho^c_i(\obs) = \pi^c_{n(i)}x_i + \Pi^c_{n(i)}(\obs)(X_i(\obs) - x_i) & \quad \forall i \in \set{P}. \tag{$\set{R}^c$}
\end{align}
\end{definition}
The day-ahead decision $x_i$ of an SP problem is hedged against uncertainty in real-time. Therefore, the first term captures the payments for providing the well-hedged solutions that are rectified in the second term by $\Pi^c_{n(i)}(\obs)x_i$. The payment corresponding to the second-stage quantity (i.e., $\Pi^c_{n(i)}(\obs)X_i(\obs)$) can then be viewed purely as a spot-market trade. The following result is well-known regarding the stochastic clearing problem in the canonical form.
\begin{proposition}\label{prop:canonicalProperties}
    Let the incremental bids satisfy $\delta^+_i, \delta^-_i > 0, \forall i \in \set{P}$ and the support of $\rvset$ is finite. The optimal solutions of the canonical stochastic clearing model result in revenue adequacy for the ISO and cost recovery for all generators in expectation under pricing-mechanism \eqref{eq:Pricing_mechanism_C}. Furthermore, the expected price distortion at node $n \in \set{N}$ satisfies $\max_{i \in \set{P}(n)}\{-\delta^+_i\}\leq~\expect{\mathbb{M}_n^c(\rv)}{}~\leq \min_{i \in \set{P}(n)}\{\delta^-_i\}$.
\end{proposition}
Notice the two critical features of the above pricing mechanism. Firstly, all elements necessary to compute the payment in \eqref{eq:Pricing_mechanism_C} can be obtained by solving extensive scenario formulation of \eqref{eq:scp}. Alternatively, they can be extracted using the solutions reported from decomposition algorithms such as the L-shaped method applied to \eqref{eq:scp_twoStage}. Secondly, the properties relating to revenue adequacy, cost recovery, and price distortion hold only in expectation. 

It is also worthwhile to notice that the optimal solutions to the clairvoyant problem can also be used with pricing-mechanism \eqref{eq:Pricing_mechanism_C}. Since these  problems are solved with complete knowledge of a scenario, it is not surprising that the resulting prices achieve cost recovery for all generators and revenue adequacy for the ISO under the corresponding scenario. However, the solutions obtained from these scenario-specific clairvoyant problems are neither implementable (as they violate the nonanticipativity requirement) nor suitable for designing day-ahead prices. Nevertheless, the canonical formulation provides an important perspective for designing pricing mechanisms; namely, capturing specific deviations and the effect of enforcing nonanticipativity may provide a path toward more meaningful pricing under uncertainty. 

Due to the decision structure of a two-stage setting, the first-stage here-and-now decisions undergo a scenario-dependent correction in the second stage. Unfortunately, the clairvoyant problem corresponding to a particular scenario, say $\obs$ is unaware of the corrections that might become necessary under a different scenario $\obs^\prime \neq \obs$. While the canonical form overcomes this deficit, this form in \ref{eq:scp} or the intrinsic first-stage form \eqref{eq:scp_twoStage}, is not directly amenable to construct a dual optimization problem without imposing finiteness on $\rvset$. The pricing mechanism \eqref{eq:Pricing_mechanism_C} fails to capture a generator's flexibility to undergo the correction from day-ahead to real-time. In other words, the payments received by generators depend only on the optimal dual solutions corresponding to the flow balance equations (possibly viewed as the stochastic location marginal prices). They do not account for the generator's flexibility to offer second-stage/real-time corrections, as the canonical form cannot capture this deviation explicitly. In the following, we present SP formulations that explicitly capture the flexibility offered by market participants in building prices while retaining optimal dispatch and computability.

\subsection{Mean-vector Formulation} \label{subsec:meanVectorModel}
To overcome the deficit of the canonical model, we develop two alternative formulations of the stochastic electricity clearing problem that are suitable to construct dual representations even when $\rvset$ has continuous support. We will use the solutions of these primal and dual problems to design the new pricing mechanisms. For this purpose, we define an extended real-valued function $\phi(\y_1, \obs) = f_1(\y_1) + H(\y_1, \obs)$ whenever $g(\y_1) = 0$ and $\y_1 \in \set{C}_1$. Otherwise, $\phi(\y_1, \obs) = +\infty$. Using the function $\phi$, the SP problem can be restated as the following:
\begin{align}\label{eq:scp_scenario}
	\min_{\y_1 \in \set{L}^\infty_{n_1}}~\{\expect{\phi(\y_1,\rv, u_1)}{\mathbb{P}} ~|~ \y_1 \in \set{N}_\infty\}.
\end{align}
The above program aims to find a $\y_1 \in \set{L}_{n_1}^\infty$ that minimizes the expectation-valued objective function while restricting the decisions to a linear subspace $\set{N}_\infty \subset \set{L}_{n_1}^\infty$. The set $\set{N}_\infty$ consists of elements that are almost surely constant and is referred to in SP literature as the set of nonanticipative elements. It captures the requirement that decision $\y_1$ must only depend on the information available at the time of decision and not on any realization of the random vector $\rv$. Our treatment of the nonanticipativity constraints and the development of the dual problems are motivated by the studies on stochastic convex programming by \cite{rockafellar1976stochastic, rockafellar1976nonanticipativity} and \cite{higle2006multistage}.

The nonanticipative set can be represented in one of two forms. We obtain the first representation of the nonanticipative set by setting $\y_1(\rv)$ to its expectation, almost surely. That is, $\set{N}_\infty = \{\y_1 \in \set{L}_{n_1}^\infty~|~ \y_1(\rv) - \expect{\y_1(\rv)}{\mathbb{P}} = 0 ~ a.s.\}$. Using this representation, we write the \emph{mean-vector formulation} of the SP problem as
\begin{align}
	\min_{\y_1 \in \set{L}_{n_1}^\infty}~& \expect{\phi(\y_1(\rv),\rv)}{\mathbb{P}} \tag{\bf P-MV} \label{eq:meanVectorPrimal}\\
	\text{subject to}&~ \y_1(\rv) - \expect{\y_1(\rv)}{\mathbb{P}} = 0,~ a.s. \label{eq:meanVector_na}
\end{align}
The detailed mean-vector formulation of the market clearing problem is presented in the online supplement \ref{subsec:appendix_meanvector}. The following result identifies a dual problem to \eqref{eq:meanVectorPrimal}. 
\begin{proposition} \label{prop:meanVectorDual} Let \eqref{eq:meanVectorPrimal} satisfy Assumption \ref{assum:SCP} and possess an optimal solution $\bar{\y}_1$ with optimal value $\nu_p^\star < \infty$. The program
\begin{align*} \label{eq:meanVectorDual}
   \sup_{\mu \in \set{L}_{n_1}^1} \expect{\phi(\y_1(\rv), \rv) + \inner{\mu(\rv) - \expect{\mu(\rv)}{\mathbb{P}}, \y_1(\rv)}}{\mathbb{P}}. \tag{\bf D-MV}
\end{align*}
has a nonempty feasible region, and its optimal value equals $\nu_p^\star$.
\end{proposition}
In the above dual optimization problem, $\mu(\obs)$ can be viewed as the dual multiplier corresponding to the nonanticipative constraint \eqref{eq:meanVector_na} associated with observation $\obs$. Defining a stochastic Lagrangian function $\mathbb{L}(\mu, \obs) = \sup_{\y_1 \in \RR^{n_1}} \{\inner{\mu(\obs) - \expect{\mu(\rv)}{\mathbb{P}}, \y_1} - \phi(\y_1, \obs)\}$, the dual problem can also be stated as $\sup_{\mu \in \set{L}_{n_1}^1} \{-\expect{L(\mu(\rv), \rv)}{\mathbb{P}}\}$. 

The primal mean-vector formulation in \eqref{eq:meanVectorPrimal} is the basis for the scenario decomposition-based methods, specifically the Progressive Hedging (PH) algorithm \citep{rockafellar1991scenarios}. Hence, a scenario decomposition algorithm can recover the desired duals computationally efficiently.

\subsubsection{Pricing Mechanism 2}
We derive the second payment mechanism from the mean-vector formulation \eqref{eq:meanVectorPrimal}. Similar to \eqref{eq:Pricing_mechanism_C}, the second mechanism also includes a day-ahead component and a real-time component which are computed using the optimal dual solutions of the respective flow-balance equations. However, unlike the previous mechanism, we include a scenario-wise adjustment for the first-stage component. Using the optimal dual $\mu_i^x$ corresponding to the nonanticipative constraint of the form in \eqref{eq:meanVector_na} for the variable $x_i(\obs)$, we define this payment mechanism as follows:
\begin{definition}
Under scenario $\obs \in \rvset$, the market participants receive a payment as follows:
\begin{align}\label{eq:Pricing_mechanism_MV}
    \rho^m_i(\obs) =( \pi_{n(i)}(\obs) + \mu^x_i(\obs))x_i(\obs) + \Pi_{n(i)}(\obs)(X_i(\obs) - x_i(\obs)) &~~~ \forall i \in \set{P}.\tag{$\set{R}^m$}
\end{align}
\end{definition}
Similar to the previous pricing mechanism, if $\rho^m_i(\obs) \leq \varphi_i^m(\obs)$, the participant $i$ achieves cost recovery, and if $\sum_{i \in \set{P}}\rho^m_i(\obs) \geq 0$ the ISO achieves revenue adequacy. The following result captures the properties of \eqref{eq:Pricing_mechanism_MV}.
\begin{theorem} \label{thm:meanVector_properties}
Let $\delta^+_i, \delta^-_i > 0 ~~ \forall i \in \set{P}$. If the optimal solutions obtained from the mean-vector formulation of the stochastic market clearing problem satisfy $\sum_{i \in \set{P}}\expect{\mu_i^{x}(\rv)}{}\expect{x(\rv)}{} \leq 0$, then pricing-mechanism \eqref{eq:Pricing_mechanism_MV} yields revenue adequacy in expectation for the ISO. Furthermore, if generator $i \in \set{G}$ satisfies $\expect{\mu_i^{x}(\rv)}{} \geq 0$, then it achieves cost recovery in expectation. Finally, the scenario-specific price distortion defined as $\mathbb{M}^m_i(\obs) := \pi_{n(i)}(\obs) + \mu_i^x(\obs) - \Pi_{n(i)}(\obs)$ satisfies
\begin{align}
    \mathbb{M}^m_i(\obs)  \in [-\delta^+_i, \delta^-_i]  + p(\obs)\mu_i^x(\obs), \quad \forall i \in \set{P}.
\end{align}
\end{theorem}
A few remarks about \eqref{eq:Pricing_mechanism_MV} are in order. Firstly, unlike \eqref{eq:Pricing_mechanism_C} where the prices and distortion $\mathbb{M}_n^c$ are defined for nodes in the power network, the prices and distortion $\mathbb{M}^m_i$ under \eqref{eq:Pricing_mechanism_MV} are defined for participants. This is due to the inclusion of participant-specific nonanticipative dual in price construction. Secondly, if there is a deviation in the cleared quantities $(X_i(\obs) - x_i(\obs) \ne 0)$, the price distortion under scenario $\obs$ is bounded by a quantity that depends on the real-time premiums $(\delta^+_i, \delta^-_i)$ and the nonanticipativity dual $(\mu^x_i(\obs))$. It is worthwhile to note that the width of the interval ($\delta_i^- + \delta_i^+$) remains the same for all scenarios, while we observe a shift equal to probability-weighted nonanticipativity dual value.

Thirdly, while the mean-vector formulation is amenable to a computationally efficient solution approach, the dual solutions obtained from solving this form provide the desired pricing properties only when certain conditions are satisfied and only in expectation. Nevertheless, the conditions depend on day-ahead quantities and nonanticipative duals, not real-time quantities.  Therefore, they can be computed upfront. Moreover, the price construction under \eqref{eq:Pricing_mechanism_MV} allows the support of the underlying stochastic process to be continuous (but bounded). We overcome these deficiencies of \eqref{eq:Pricing_mechanism_MV} in the next pricing mechanism.

\subsection{State-vector Formulation} \label{subsec:stateVectorModel}
The second formulation utilizes an alternative way to express the nonanticipativity set $\set{N}_\infty$. This form involves a state vector $\chi_1 \in \RR^{n_1}$ and is given by $\set{N}_\infty = \{\y_1 \in \set{L}_{n_1}^\infty ~|~\y_1(\rv) - \chi_1 = 0,~a.s.\}$. Recall that $\phi(\y_1, \obs)$ takes a finite value $f_1(\y_1) + H(\y_1, \obs)$ only for feasible first-stage decisions, i.e., when $g(\y_1) = 0$ and $\y_1 \in \set{C}_1$, and $+\infty$ otherwise. Using the alternative representation of nonanticipativity requirement, the \emph{state-vector formulation} of the  SP problem is given by
\begin{align} \label{eq:stateVectorPrimal}
	\min_{\y_1 \in \set{L}_{n_1}^\infty, \chi_1 \in \RR^{n_1}}~& \expect{\phi(\y_1(\rv),\rv)}{\mathbb{P}} \tag{\bf P-SV}\\
	\text{subject to}&~ \y_1(\rv) - \chi_1 = 0,~ a.s. \label{eq:stateVector_na}
\end{align} 
Notice that the above is an optimization problem in an extended space $\set{L}_{n_1}^\infty \times \RR^{n_1}$ obtained by the inclusion of the state vector $\chi_1 \in \RR^{n_1}$. We denote its feasible region by $\widehat{\set{N}}_\infty$. The next result identifies a suitable dual to the above \eqref{eq:stateVectorPrimal}. For this purpose, we introduce a set of multipliers $\set{M} = \{\sigma \in \set{L}_{n_1}^1~|~ \expect{\sigma(\rv)}{\mathbb{P}} = 0\}$ and the conjugate function $\phi^*(\sigma,\obs) = \sup_{\y_1 \in \RR^{n_1}} \{\inner{\sigma, \y_1} - \phi(\y_1, \obs)\}$.
\begin{proposition} \label{prop:stateVectorDual} Let \eqref{eq:stateVectorPrimal} satisfy Assumption \ref{assum:SCP} and possess an optimal solution $(\bar{\y}_1, \bar{\chi}_1)$ with optimal value $\nu_p^\star < \infty$. The program
\begin{align}\label{eq:stateVectorDual}
	\sup_{\sigma \in \set{L}^1_{n_1}} \big\{-\expect{\phi^\star(\sigma(\rv), \rv)}{\mathbb{P}} ~|~ \expect{\sigma(\rv)}{\mathbb{P}} = 0\big\} \tag{\bf D-SV}
\end{align}
has a nonempty feasible region, and its optimal value equals $\nu_p^\star$.
\end{proposition}
To establish the above result, we utilize tools from the theory of conjugate duality of convex analysis \citep{rockafellar1974conjugate}. The dual program presented above exhibits key features we exploit in developing the pricing mechanisms. First, the nonanticipative duals $\sigma \in \set{M}$ can be viewed as equilibrium prices. We establish this fact formally in the following result.
\begin{proposition} \label{prop:infoPrice}
Consider a Lagrangian $\mathbb{L}:\set{L}_{n_1}^\infty \times \RR^{n_1} \times \set{L}_{n_1}^1 \rightarrow \RR$ associated with the perturbed problem defined as
\begin{align}
	\mathbb{L}(\y_1, \chi_1, \sigma) = \expect{\phi(\y_1(\rv), \rv) + \inner{\y_1(\rv) - \chi_1, \sigma(\rv)}}{\mathbb{P}}.
\end{align}
A solution $(\bar{\y}_1, \bar{\chi}_1) \in \widehat{\set{N}}_\infty$ is an optimal solution of \eqref{eq:stateVectorPrimal} if and only if there exists a multiplier vector $\bar{\sigma} \in \set{M}$ such that 
\begin{align*}
 	(\bar{\y}_1, \bar{\chi}_1) \in \argmin_{(\y_1, \chi_1) \in \set{L}_{n_1}^\infty \times \RR^{n_1}} \mathbb{L}(\y_1, \chi_1, \bar{\sigma}).
\end{align*}
\end{proposition}
The above result shows that $(\bar{\y}_1, \bar{\chi}_1, \bar{\sigma})$ is a saddle point of the Lagrangian $\mathbb{L}$. For $v \in \set{L}_{n_1}^\infty$, let us define a function $F: \set{L}_{n_1}^\infty \times \RR^{n_1} \times \set{L}_{n_1}^\infty \rightarrow (-\infty, \infty]$ as
\begin{align*}     
F(\y_1, \chi_1, v) =  \begin{cases} 
    f_1(\y_1) + \expect{H(\y_1, \rv)}{\mathbb{P}} & \text{if } \y_1 \in \set{C}_1, g_1(\y_1) = 0, \y_1(\rv) - \chi_1 = v(\rv) \\
    +\infty & \text{otherwise}.
\end{cases}
\end{align*}
The function $F$ has been referred to in the literature as the ``bivariate function'' \citep{bauschke2011convex}, the ``perturbation function'' \citep{zalinescu2002convex}, and more recently, as the \emph{Rockafellian} \citep{royset2021good}. Notice that the SP problem in \eqref{eq:stateVectorPrimal} is equivalent to finding $(\y_1, \chi_1)$ that minimizes the Rockafellian $F(\y_1, \chi_1, 0)$. Let $\varphi(v) = \inf_{\y_1, \chi_1} F(\y_1, \chi_1, v)$, then
\begin{align}
    \varphi(0) = \mathbb{L}(\bar{\y}_1, \bar{\chi}_1, \bar{\sigma})  =~& \inf_{(\y_1, \chi_1)} \mathbb{L}(\y_1, \chi_1, \bar{\sigma}) \notag\\
    \leq~& \inf_{(\y_1, \chi_1)}~ \bigg\{ \inf_{v \in \set{L}_{n_1}^1} F(\y_1, \chi_1, v) + \expect{\inner{v(\rv), \bar{\sigma}(\rv)}}{\mathbb{P}} \bigg\} \notag\\
    \leq~& \inf_{(\y_1, \chi_1)} F(\y_1, \chi_1, v) + \expect{\inner{v(\rv), \bar{\sigma}(\rv)}}{\mathbb{P}}  \notag\\
    \Rightarrow \qquad \varphi(0) \leq~& \varphi(v) + \expect{\inner{v(\rv), \bar{\sigma}(\rv)}}{\mathbb{P}}.
\end{align}
The above inequality shows that the deviation $v$ results in a value of $\varphi(v)$ in place of $\varphi(0)$, however, this change is associated with a cost of $\expect{\inner{v(\rv), \bar{\sigma}(\rv)}}{\mathbb{P}}$. It is worthwhile to note that $\expect{\inner{v(\rv), \bar{\sigma}(\rv)}}{\mathbb{P}}$ is only the perceived cost at the time the first-stage decision is taken.  Therefore, $\bar{\sigma}$ must be interpreted as a system of \emph{equilibrium prices}. These prices indicate that there is no incentive to deviate from the first-stage implementable decision $\bar{\chi}_1$. Since the decision $\bar{\chi}_1$ is chosen for implementation in a  nonanticipative manner, it is determined before the realization of uncertainty. If a specific scenario $\obs$ is realized and we could change the decision to $\bar{\y}_1(\obs) = \bar{\chi}_1(\obs) + v(\obs)$ using the complete knowledge of the realized scenario, then the cost of doing so is $\bar{\sigma}(\obs)v(\obs)$. Therefore, $\bar{\sigma}(\obs)$ can also be viewed as the \emph{price of information} realized under scenario $\obs$. 

Second, it can be seen that \eqref{eq:stateVectorPrimal} is equivalent to \eqref{eq:scp} and \eqref{eq:meanVectorPrimal} obtained by setting $\chi_1 = \expect{\y_1(\rv)}{\mathbb{P}}$. The dual problems \eqref{eq:meanVectorDual} and  \eqref{eq:stateVectorDual} are also equivalent. This equivalence is established by simply noting that $\sigma(\rv) = \mu(\rv) - \expect{\mu(\rv)}{\mathbb{P}}$, almost surely, is feasible to \eqref{eq:stateVectorDual}. A dual of \eqref{eq:scp} can be obtained only in the case of finite support for $\rvset$. We summarize the relationship between the solutions of all three alternative SP formulations for the case when $\rvset$ has finite support in the following result.  
\begin{proposition} Let $(\pi^c, (\Pi^c(\obs))_{\forall \obs \in \rvset})$ denote the optimal dual solutions corresponding to flow-balance equations, \eqref{eq:da_flowBalance_original} and \eqref{eq:rt_flowBalance} in \eqref{eq:scp_twoStage}. Similarly, let $(\pi^m(\obs), \Pi^m(\obs), \mu(\obs))_{\forall \obs \in \rvset}$ denote the optimal multipliers for \eqref{eq:da_flowBalance_original}, \eqref{eq:rt_flowBalance}, and nonanticipativity constraint \eqref{eq:meanVector_na} in \eqref{eq:meanVectorPrimal}. Finally, let $(\pi^s(\obs), \Pi^s(\obs), \sigma(\obs))_{\forall \obs \in \rvset}$ denote the optimal multipliers for \eqref{eq:da_flowBalance_original}, \eqref{eq:rt_flowBalance}, and nonanticipativity constraint \eqref{eq:stateVector_na} in \eqref{eq:stateVectorPrimal}. These solutions satisfy 
    \begin{enumerate}[label=\roman*.]
        \item $\pi^c = \mathbb{E}[\pi^{m}(\obs)] = \mathbb{E}[\pi^{s}(\obs)]$; \label{thm:solutionRelationship_DA_duals}
        \item $\mathbb{E}[\sigma(\rv)] = 0$; \label{thm:solutionRelationship_NA_duals}
        \item If $\pi^s(\obs) = \pi^m(\obs)$, then $\sigma(\obs) = \mu(\obs) - \expect{\mu(\rv)}{}$ for all $\obs \in \Omega$; \label{thm:solutionRelationshop_MV_SV}
	    \item $\Pi^c(\obs) = \Pi^{m}(\obs) = \Pi^{s}(\obs)$, for all $\obs \in \Omega$. \label{thm:solutionRelationship_RT_duals}
    \end{enumerate} \label{thm:solutionRelationship}
\end{proposition}
It is worthwhile to note the following about the above theorem. The dual optimal solutions corresponding to the day-ahead flow-balance equations of the mean-vector and state-vector form are equal only in expectation to the dual optimal solution corresponding to the day-ahead flow-balance equation of the canonical form. On the other hand, those corresponding to the real-time flow-balance equations are equal for every observation. Finally, the relationship in item $\ref{thm:solutionRelationshop_MV_SV}$ enables us to use the scenario decomposition methods such as progressive hedging as a solution approach and translate the resulting dual solution into those corresponding to \eqref{eq:stateVectorPrimal}. In this sense, we retain an efficient computational approach to solving stochastic clearing problems.

\subsubsection{Pricing Mechanism 3}
We derive the third payment mechanism from the state-vector formulation \eqref{eq:stateVectorPrimal}. Similar to previous payment mechanisms, this mechanism also includes a day-head component and a real-time component which are computed using the optimal dual solutions of the respective flow-balance equations. Furthermore, like in \eqref{eq:Pricing_mechanism_MV}, we include a scenario-wise adjustment for the first-stage component. We define this payment mechanism as follows:
\begin{definition}
Under scenario $\obs \in \rvset$, the market participants receive a payment given by
\begin{align}\label{eq:Pricing_mechanism_SV}
    \rho^s_i(\obs) = (\pi^s_{n(i)}(\obs) + \sigma^x_i(\obs))x_i(\obs) + \Pi^s_{n(i)}(\obs)(X_i(\obs) - x_i(\obs)) \quad \forall i \in \set{P}. \tag{$\set{R}^s$}
\end{align}
Here, $\sigma_i^x(\obs)$ is the dual corresponding to the nonanticipative constraint of the form in \eqref{eq:stateVector_na} for the variable $x_i(\obs)$.
\end{definition}

Notice that the payment mechanism differs from \eqref{eq:Pricing_mechanism_MV} only in its use of $\sigma_i^x(\obs)$ instead of $\mu_i^x(\obs)$. Furthermore, the day-ahead component in this pricing mechanism is scenario-dependent that satisfies $\expect{(\pi^s_{n(i)}(\rv) + \sigma^x_i(\rv)}{} = \pi^c_{n(i)}$ (due to Theorem \ref{thm:solutionRelationship}, parts $\ref{thm:solutionRelationship_DA_duals}$ and $\ref{thm:solutionRelationship_NA_duals}$). In other words, the day-ahead component of $\eqref{eq:Pricing_mechanism_MV}$ aligns in expectation with the day-ahead component of $\eqref{eq:Pricing_mechanism_C}$. The clairvoyant problem can identify the best dispatch plan for a given scenario. With the aim of hedging against uncertainty, the SP optimal dispatch deviates from the best scenario-specific dispatch. The quantity $\sigma^x_i(\obs)x_i(\obs)$ included in \eqref{eq:Pricing_mechanism_SV} allows us to reflect the value offered by a participant $i$ to attain a well-hedged stochastic solution in their payment. For this pricing mechanism, the scenario-specific price distortion is given by $\mathbb{M}_i^s(\obs) := \pi^s_{n}(\obs)  + \sigma^x_i(\obs) - \Pi^s_{n}(\obs),~ \forall i \in \set{P}, \obs \in \Omega$. As in \ref{eq:Pricing_mechanism_MV}, notice that the prices and price distortion under \eqref{eq:Pricing_mechanism_SV} are defined for participants. In other words, multiple participants at the same node in the power network can receive different payments and observe different distortions based on the value of their respective nonanticipative dual. This pricing mechanism exhibits the following characteristics. 

\begin{theorem} \label{thm:stateVector_properties}
    Under the assumption that $\delta^+_i, \delta^-_i > 0 ~~ \forall i \in \set{P}$, the payment mechanism \eqref{eq:Pricing_mechanism_SV}, computed using the optimal solutions of the \eqref{eq:stateVectorPrimal}, yield revenue adequacy in expectation for the ISO and cost recovery under every scenario for all generators. Furthermore, the scenario-specific price distortion satisfies $$-\delta_i^+ \leq \mathbb{M}_{i}^s(\obs) \leq \delta_i^-\qquad \forall i \in \set{P}, \obs \in \Omega.$$
\end{theorem}

The above results demonstrate that the pricing-mechanism \eqref{eq:Pricing_mechanism_SV} retains the desired properties of revenue adequacy in expectation from \eqref{eq:Pricing_mechanism_MV}. However, pricing-mechanism \eqref{eq:Pricing_mechanism_SV} provides a stronger guarantee of scenario-specific cost recovery for all participating generators. Furthermore, under pricing-mechanism \eqref{eq:Pricing_mechanism_C}, we attain price consistency in the sense that the expected price distortion is bounded to an interval that depends only on $\delta_i+$ and $\delta_i^-$ (see Proposition \ref{prop:canonicalProperties}). This result is also strengthened under pricing-mechanism \eqref{eq:Pricing_mechanism_SV} as we attain bounds on scenario-specific price distortions. Together with the fact that $\expect{\sigma_i^x(\rv)}{} = 0$ for all $i \in \set{P}$ (Proposition \ref{thm:solutionRelationship}, Part $\ref{thm:solutionRelationship_NA_duals}$), the bounds on scenario-specific price distortions imply bounds on expected price distortion. The premise of the stochastic clearing problem is that the real-time adjustments to quantities cleared in the day-ahead market incur a premium. While keeping the day-ahead and real-time prices consistent under every scenario is desirable (i.e., $\mathbb{M}_i^s(\rv) = 0$, almost surely), such an endeavor is impossible as long as the adjustment costs ($\delta_i^+$ and $\delta_i^-$) are nonzero. This also implies that it is impossible to achieve revenue adequacy under every scenario while maintaining the optimality of dispatch. Therefore, the above result establishes the best one could achieve under a stochastic clearing model. Finally, we argue that scenario-specific cost recovery and bounded price distortions encourage generators to participate in the market. On the other hand, it suffices for the ISO to be revenue adequate in the long run, a possibility under a pricing mechanism that guarantees this property in expectation.  

Before we close this section, we emphasize that the primal and dual formulations built based on the mean-vector and state-vector representations of nonanticipativity are equivalent. However, the interpretation of the nonanticipative dual and its role in price formation is significantly different. The duals from the mean-vector formulation result (Theorem \ref{thm:meanVector_properties}) in revenue adequacy and cost recovery in expectation only when certain additional conditions are met. On the other hand, the duals from the state-vector formulation ensure revenue adequacy in expectation and scenario-specific cost recovery as shown in Theorem \ref{thm:stateVector_properties}. Therefore, \ref{eq:Pricing_mechanism_SV} is a suitable pricing mechanism for the stochastic electricity market clearing problem. We illustrate these results numerically in the next section. 

\section{Numerical Illustrations} \label{sec:expriments}
This section illustrates the properties of all the pricing mechanisms presented in \S\ref{sec:pricing_mechanisms_analysis}. We implemented all the models using C++ Concert technology and solved them using CPLEX 12.9 solver. The experiments were conducted on an Intel Core i3, 2.20 GHz processor, and 8GB RAM. 

For our experiments, we utilize three test systems. The first two test systems are those used in \cite{pritchard2010single} and \cite{zavala2017stochastic}. We refer to these systems as PZP-6 and ZKAB-6, respectively. Instances of the first two test systems have $25$ scenarios. We use the IEEE-30 instance from \cite{UWPTest} to set up the third test system. This system has a more general network topology and utilizes wind generation scenarios generated using the change point-based model proposed in \cite{ariyarathne2022change}. We use $200$ independently generated scenarios in our experiments with this test system. The change point-based model allows us to examine the pricing mechanism under realistic wind outcomes by capturing wind speed's nonstationarity and spatiotemporal correlation. We refer to this system as SODA-30. We refer the reader to \S\ref{sec:testSystems} in the online supplement for more details about the test systems.
 
\begin{figure}[t!]
\centering
\begin{subfigure}{.5\textwidth}
  \centering
  \frame{\includegraphics[width=.9\linewidth,trim=0 0 10cm 1em,clip]{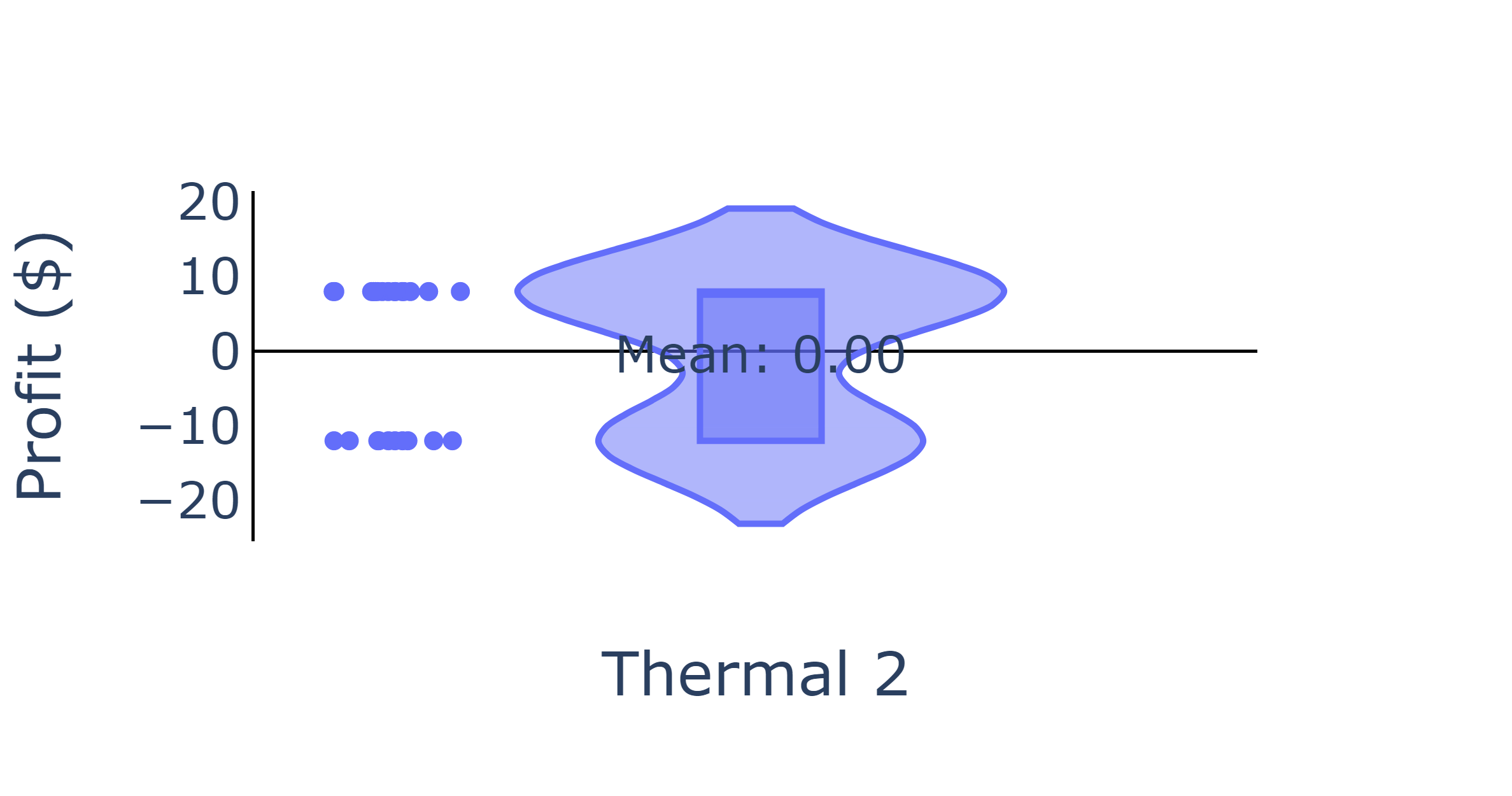}}
  \caption{Cost recovery}
  \label{fig:pritchard6_thermal2_costRecovery}
\end{subfigure}%
\begin{subfigure}{.5\textwidth}
  \centering
  \frame{\includegraphics[width=.9\linewidth,trim=0 0 10cm 1em,clip]{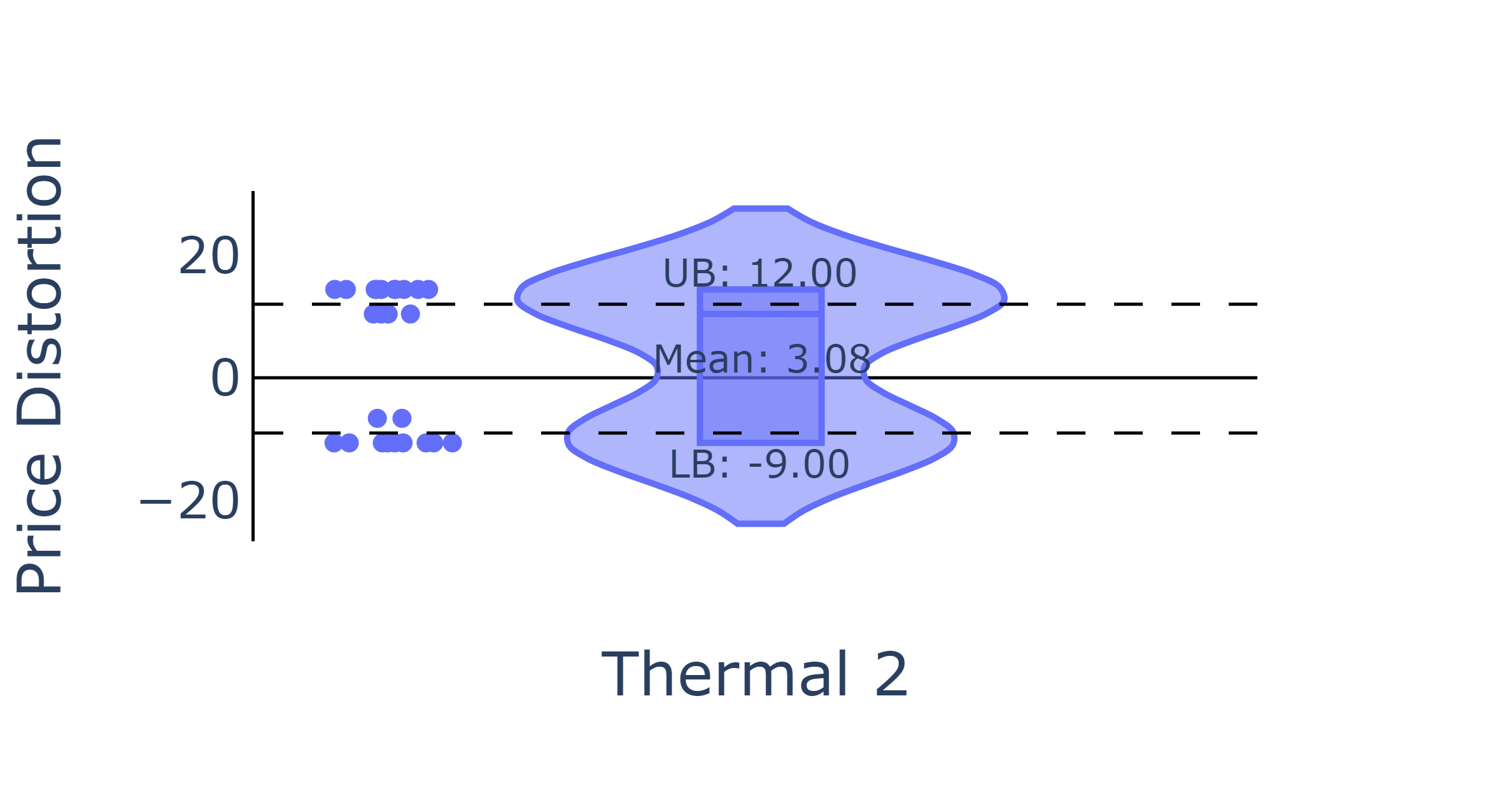}}
  \caption{Price distortion ($\delta_i^+ = 9, \delta_i^- = 12$)}
  \label{fig:pritchard6_thermal2_priceDistortion}
\end{subfigure}
\caption{Cost recovery and price distortion result for thermal 2 in the canonical formulation.}
\label{fig:pritchard6_canonical}
\end{figure}
To begin, we numerically verified the relationships between the primal and dual solutions established in Theorem \ref{thm:solutionRelationship} for all three test systems. Recall that the pricing mechanism \ref{eq:Pricing_mechanism_C} guarantees cost recovery, revenue adequacy, and price distortion properties only in expectation. The expected revenue and cost for the system operator of PZP-6 were $\$30154.1$ and $\$7067.88$, respectively. Figure \ref{fig:pritchard6_thermal2_costRecovery} illustrates profits for a thermal generator in PZP-6. Notice that while the profits are negative for some scenarios and positive for others, however, the mean value is zero. This indicates that the generator achieves cost recovery in expectation. Further, Figure \ref{fig:pritchard6_thermal2_priceDistortion} shows that there are some scenarios in which the price differences are outside their distortion interval $[-\delta_i^+, \delta_i-]$. However, in expectation, the generator archives a price distortion of $3.08 \in [-9,12]$.

\subsection{Verification of Pricing Mechanism \ref{eq:Pricing_mechanism_MV}}
Next, we focus on the solutions obtained from the mean-vector formulation and the resulting prices constructed using \ref{eq:Pricing_mechanism_MV}. Recall that, in Theorem \ref{thm:meanVector_properties}, to achieve revenue adequacy in expectation, the mean-vector solutions have to satisfy $\sum_{i \in \set{P}}\expect{\mu^{x^*}_i(\rv)}{}\expect{x^*_i(\rv)}{} \leq 0$. Table \ref{tab:revenueAdequcy} shows results verifying the revenue adequacy under \eqref{eq:Pricing_mechanism_MV}, including the condition, cost incurred (payments made to the generator), revenue generated (payments received from the loads), and the net income for the ISO. As shown in Table \ref{tab:revenueAdequcy}, PZP-6 and ZKAB-6 test systems fail to satisfy this condition. While the ISO for PZP-6 is not revenue-adequate, the ZKAB-6 ISO has a positive net income, indicating revenue adequacy in expectation. On the other hand, the SODA-30 system satisfies the condition and achieves revenue adequacy, as shown by the positive net income in Table \ref{tab:revenueAdequcy}. These results validate our claim that the condition is sufficient for revenue adequacy but not necessary.

\begin{table}[t!] \renewcommand{\arraystretch}{1.2} \centering 
	\resizebox{0.99\textwidth}{!}{\begin{tabular}{|c|c c c c || c c c|} \hline
	\multirow{2}{*}{System} & \multicolumn{4}{c||}{\ref{eq:Pricing_mechanism_MV}} & \multicolumn{3}{c|}{\ref{eq:Pricing_mechanism_SV}} \\ \cline{2-8}
	& Condition & Cost & Revenue & Net income & Cost & Revenue & Net income \\ \hline
	PZP-6  & 546,674 & \$7,067.88 & -\$509,902.24 & -\$502,834.36 & \$7,067.88 & \$22,234.08 & \$29,301.96 \\ 
	ZKAB-6 & 12,562 & \$19,482.40 & \$161,922.56 & \$181,404.96 & \$19,482.40 & \$162,200.00 & \$181,682.40\\
	SODA-30& -146.65 & \$45,957.25 & \$244,880.00 & \$290,837.25 & \$45,957.25 & \$245,929.54 & \$291,886.79 \\ \hline	
	\end{tabular}} 
	\caption{Results verifying revenue adequacy in expectation under \ref{eq:Pricing_mechanism_MV} and \ref{eq:Pricing_mechanism_SV}} \label{tab:revenueAdequcy}
\end{table}

\begin{figure}[b!]
\centering
    \frame{\includegraphics[width = 0.9\textwidth]{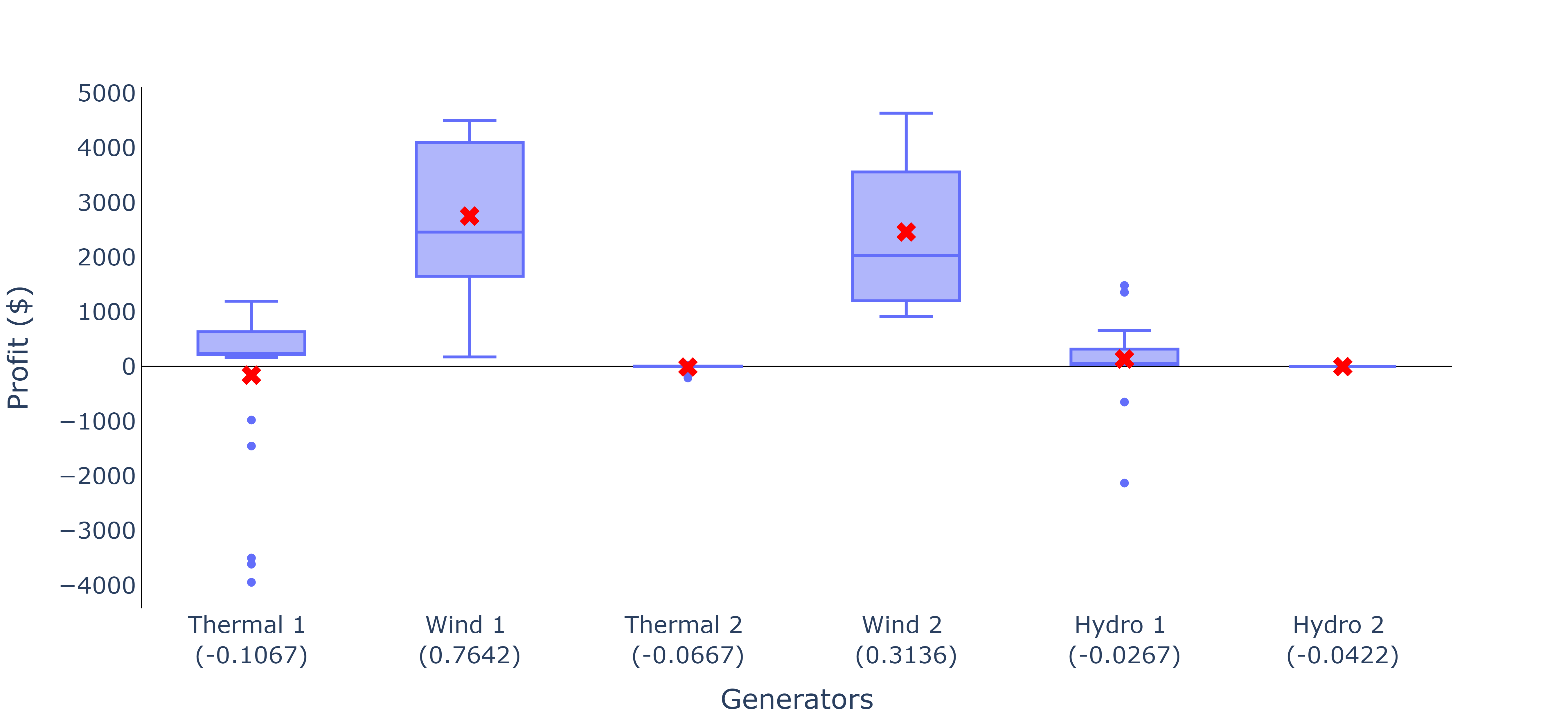}}
    \caption{Cost recovery for individual scenarios and in expectation under \ref{eq:Pricing_mechanism_MV}.}
\label{fig:pritchard6_meanVector_costRecover}
\end{figure}

Theorem \ref{thm:meanVector_properties} identifies that generators achieve cost recovery in expectation if $\expect{\mu^{x^*}_i(\rv)}{} \geq 0$. This result is illustrated in Figure \ref{fig:pritchard6_meanVector_costRecover}. The boxplots capture the scenario-specific profits for all six generators in PZP-6. The crosses inside the box plots indicate the expected profit, and the parenthetical values under the generator names are the values of $\expect{\mu^{x^*}_i(\rv)}{}$ for the respective generators.
Notice that generators with $\expect{\mu^{x^*}_i(\rv)}{} > 0$, e.g., Wind 1 and Wind 2, are revenue adequate in expectation. On the other hand, Thermal 2, Hydro 1, and Hydro 2 also have a positive expected profit even though the condition is violated. This illustrates that $\expect{\mu^{x^*}_i(\rv)}{} \geq 0$ is only a sufficient condition for cost recovery in expectation. Finally, the figure indicates negative profits for some scenarios, most prominently for Thermal 1 and Hydro 1 generators, indicating that the mean-vector formulation cannot achieve scenario-wise cost recovery for all the generators.

\subsection{Verification of Pricing Mechanism \ref{eq:Pricing_mechanism_SV}}
Table \ref{tab:revenueAdequcy} also presents the cost, revenue, and net
income for the ISO under \ref{eq:Pricing_mechanism_SV}. The results indicate that the expected total revenue generated from meeting the demand is higher than the expected total payment made to the generators. This results in revenue adequacy in expectation for the operator. We next demonstrate the property on cost recovery for all generators in every scenario.
\begin{figure}[t!]
\centering
\begin{subfigure}{\textwidth}
\centering
    \includegraphics[width = 0.9\textwidth]{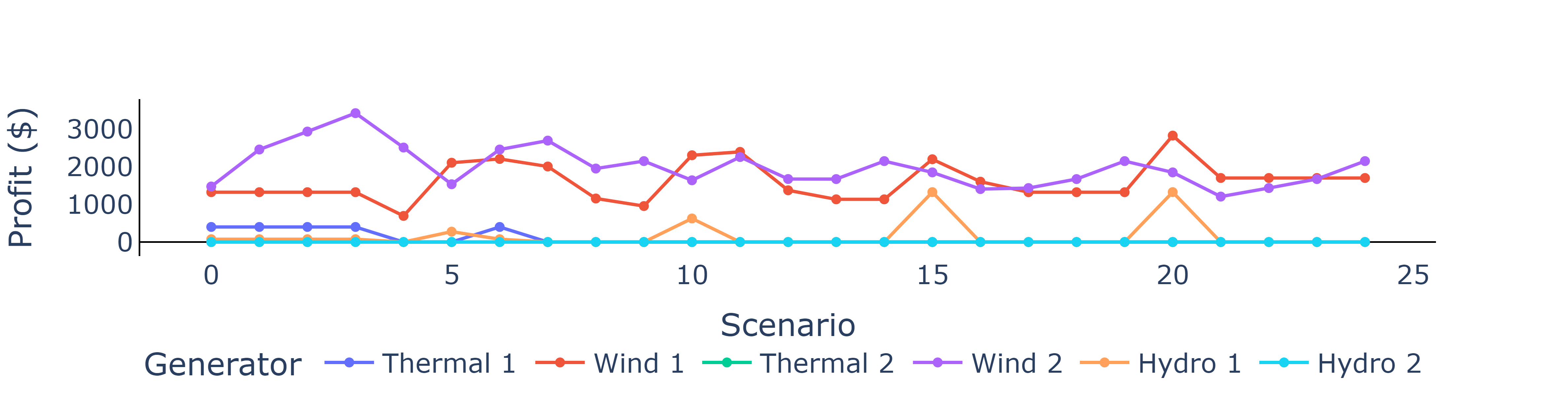}
    \caption{System: PZP-6.}
    \label{fig:pritchard6_costRecovery}
\end{subfigure}
\begin{subfigure}{\textwidth}
\centering
    \includegraphics[width = 0.9\textwidth]{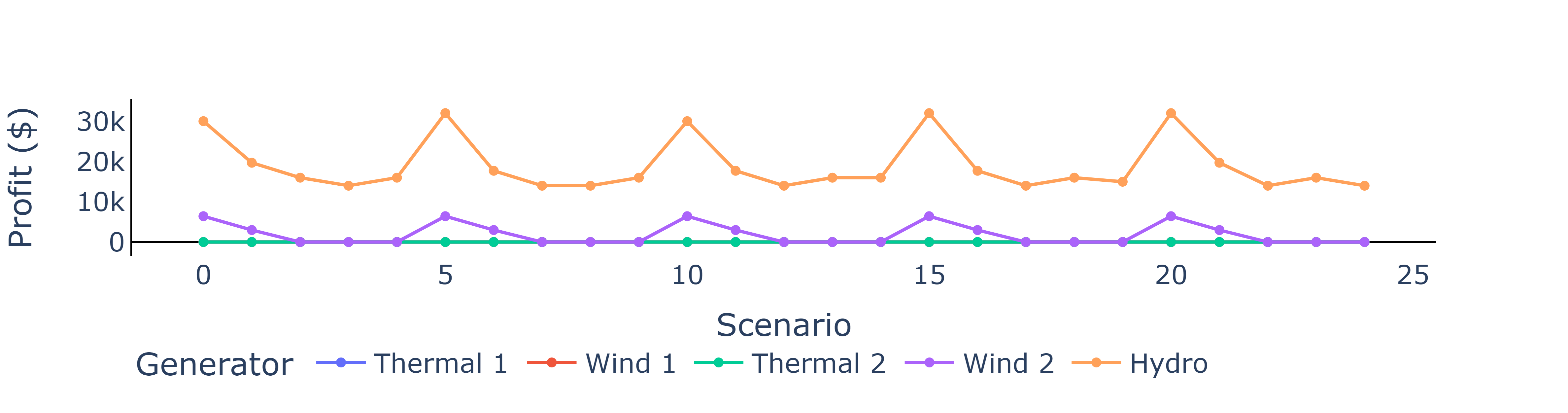}
    \caption{System: ZKAB-6.}
    \label{fig:zavala6_costRecovery}
\end{subfigure}
\begin{subfigure}{\textwidth}
\centering
    \includegraphics[width = 0.9\textwidth]{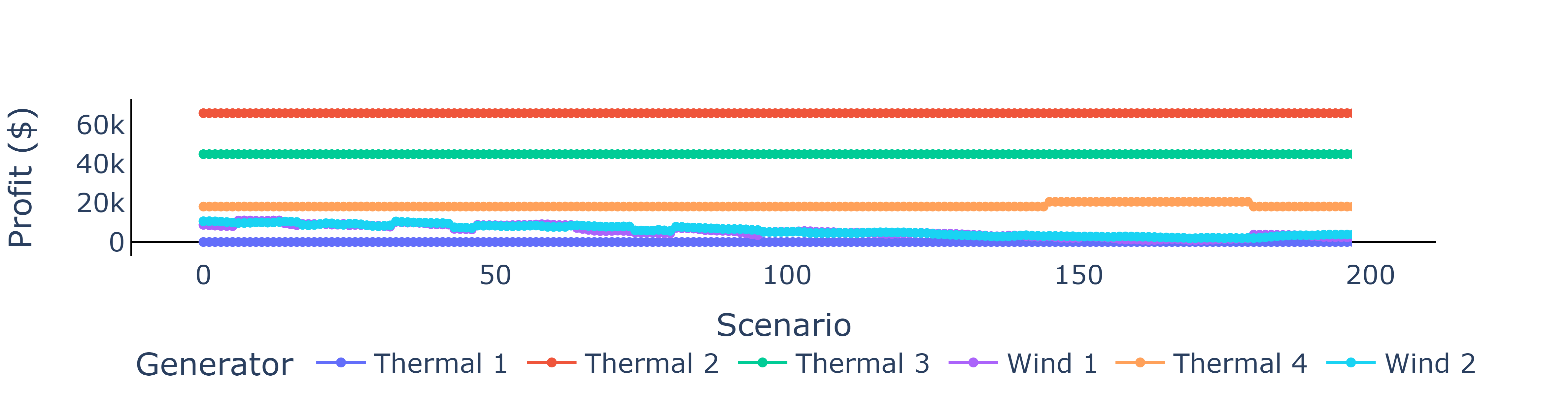}
    \caption{System: SODA-30.}
    \label{fig:soda30_costRecovery}
\end{subfigure}
\caption{State-vector cost recovery results.}
\label{fig:SV_costRecovery}
\end{figure}
Figure \ref{fig:SV_costRecovery} depicts the profit of each generator under every scenario for all three systems. Notice that all the generators recover their costs under every scenario. While the wind generators make the most profit in PZP-6, the hydro generator and Thermal-1 generate the most profits in ZKAB-6 and SODA-30, respectively. 

% In Figure \ref{fig:pritchard6_costRecovery}, the two wind generators are making significant profits compared to the other generators due to their low production costs of $\$0$. We see similar results for wind generators under pricing mechanism \ref{eq:Pricing_mechanism_C}. However, unlike \ref{eq:Pricing_mechanism_C}, pricing mechanisms \ref{eq:Pricing_mechanism_SV}, can recover their costs under every scenario. The cost recovery property also holds for conventional and hydro generators, encouraging the generators to participate in the market. Similar results are observed in ZKAB-6 and SODA-30 systems as shown in Figures \ref{fig:zavala6_costRecovery} and \ref{fig:soda30_costRecovery}, respectively. 

Finally, we discuss the price distortion results under pricing mechanism \ref{eq:Pricing_mechanism_SV}. 
\begin{figure}[t!]
\centering
\begin{subfigure}{0.32\textwidth}
\centering
    \includegraphics[width = 0.99\textwidth,trim=0 0 9cm 1em,clip]{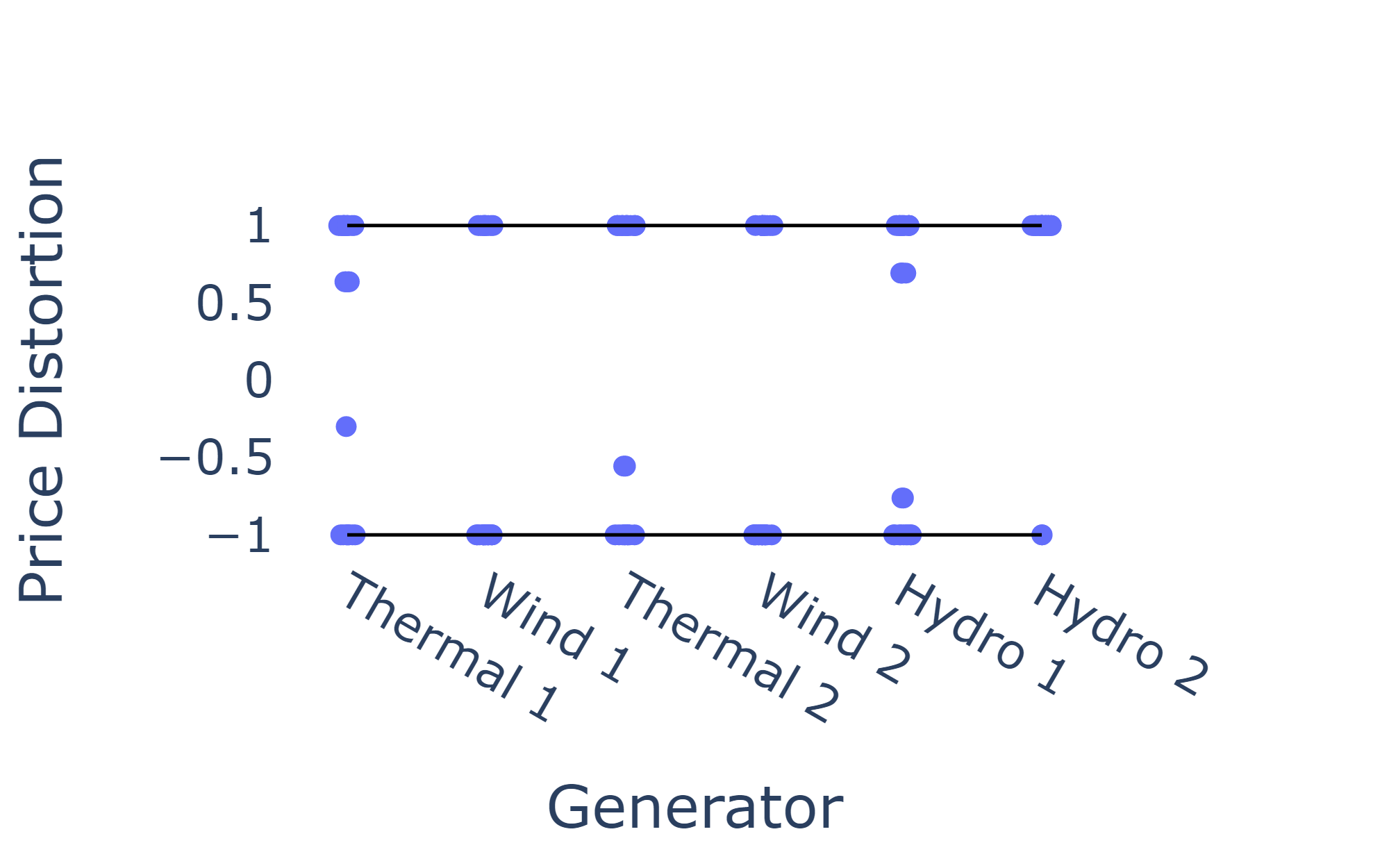}
    \caption{System: PZP-6.}
    \label{fig:pritchard6_priceDistortion}
\end{subfigure}
\begin{subfigure}{0.32\textwidth}
\centering
    \includegraphics[width = 0.99\textwidth,trim=0 0 12cm 1em,clip]{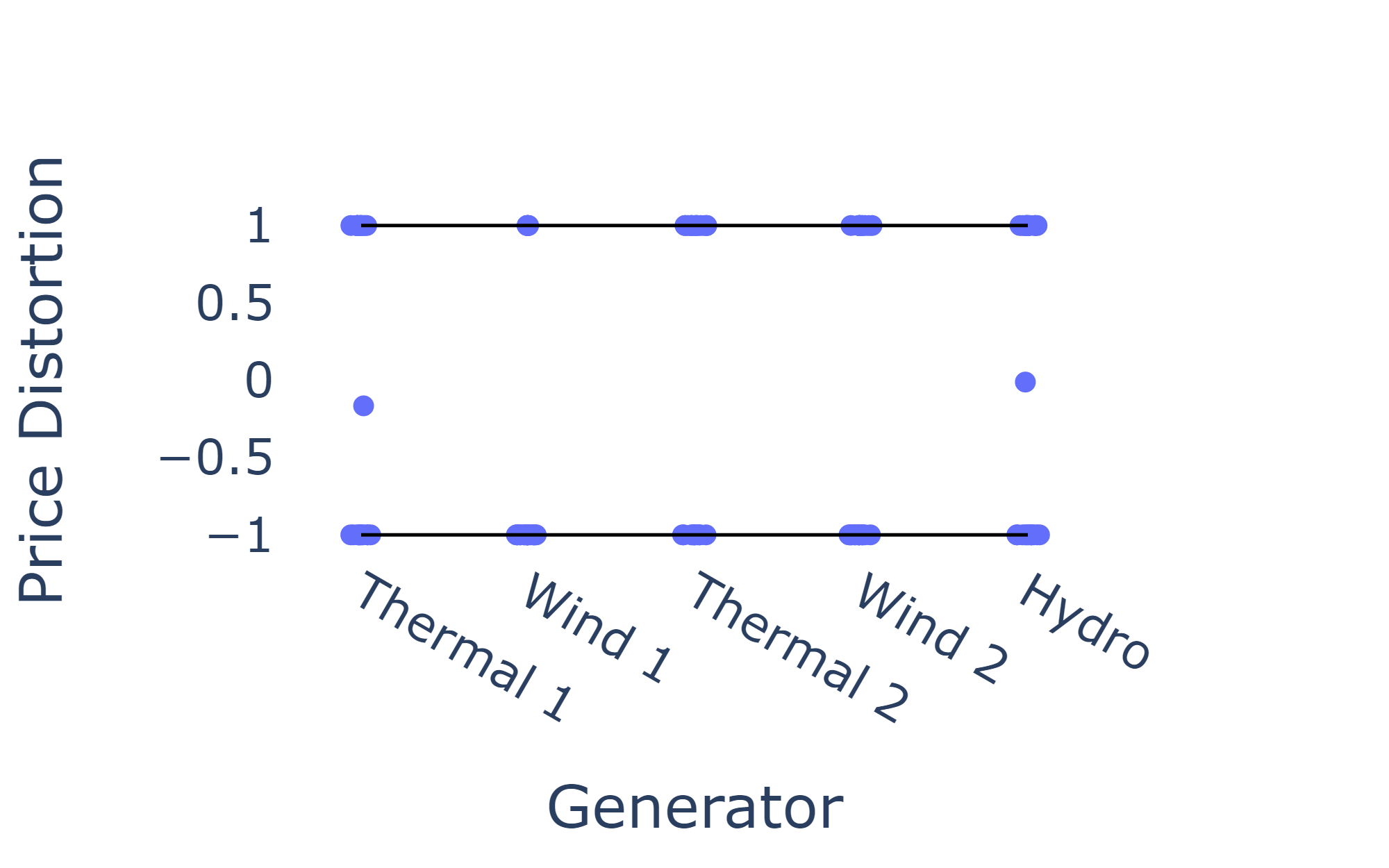}
    \caption{System: ZKAB-6.}
    \label{fig:zavala6_priceDistortion}
\end{subfigure}
\begin{subfigure}{0.32\textwidth}
\centering
    \includegraphics[width = 0.99\textwidth,trim=0 0 12cm 1em,clip]{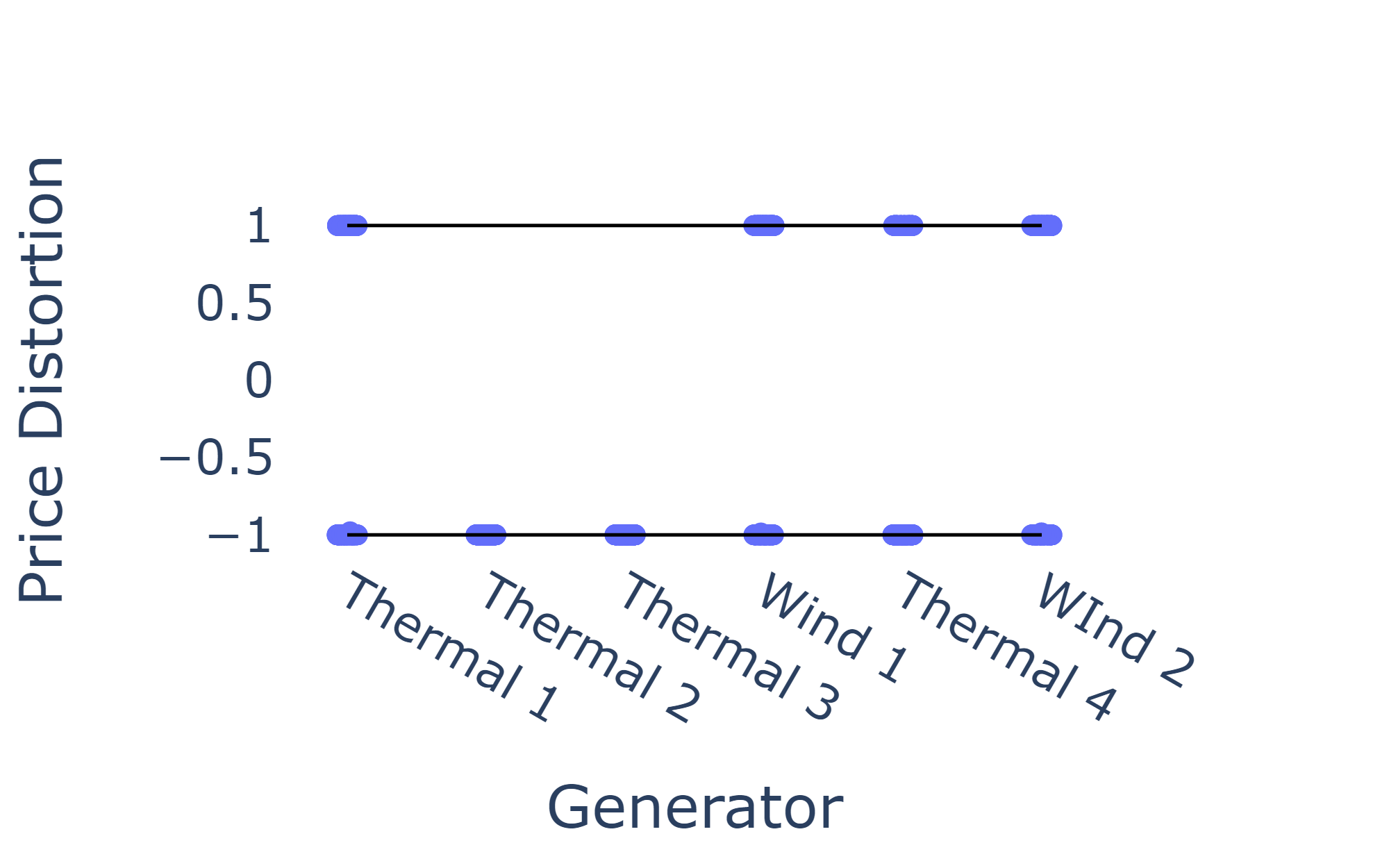}
    \caption{System: SODA-30.}
    \label{fig:soda30_priceDistortion}
\end{subfigure}
\caption{State-vector price distortion results for all three test systems.}
\label{fig:SV_priceDistortion}
\end{figure}
Recall that we define price distortion as the difference between the day-ahead and real-time unit payments. Theorem \ref{thm:stateVector_properties} bounds the price distortion to an interval defined by the deviation premiums as $[-\delta^+_i, \delta^-_i]$. Figure \ref{fig:SV_priceDistortion} depicts the price distortions for all generators in the three test systems scaled to the $[-1, 1]$. Notice that all price distortions are within their respective distortion bounds. This implies that the prices formed under \ref{eq:Pricing_mechanism_SV} do not deviate outside an interval set by the market participants' real-time offers. 

\section{Conclusion}\label{sect:conclusion}
In this paper, we presented alternative SP models, namely, mean-vector and state-vector models, for the electricity market clearing problem. While the canonical SP model can capture the stochasticity in the real-time market and is well studied, it fails to construct prices that adapt to uncertain scenarios. To address this issue, we developed electricity market clearing models that are based on the concept of nonanticipativity in SP. Modeling first-stage decision variables, in our case, day-ahead market decisions for each scenario, and tying them together using nonanticipativity constraints allowed us to extract scenario-specific dual solutions. In particular, the dual solution corresponding to the nonanticipativity constraint captured the cost of deviating from the scenario's optimal solution. 

Alternative approaches to model nonanticipativity resulted in two different models and corresponding pricing mechanisms. While the mean-vector form is more compatible with solution algorithms (notably, the progressive hedging algorithm), the pricing mechanism \ref{eq:Pricing_mechanism_MV}  based on this form yields revenue adequacy and cost recovery in expectation only when additional conditions are met. Although these conditions depend only on primal solutions and nonanticipativity duals and are easy to verify, they are challenging to interpret and hence, less desirable in practice. 

The state-vector form and the corresponding pricing mechanism \ref{eq:Pricing_mechanism_SV} overcame these deficiencies in the mean-vector form. The prices constructed under \ref{eq:Pricing_mechanism_SV} always ensure revenue adequacy for the system operator in expectation. Furthermore, \ref{eq:Pricing_mechanism_SV} offered the desirable property of cost recovery under every scenario for all the participating generators. Finally, we determined uniform bounds on scenario-wise price distortion that only depend on participants' real-time premiums. The cost recovery and bounded price distortion results provide the much-needed incentives for generators to participate in stochastic markets. 

Our approach to constructing stochastic market clearing models, nonanticipative-based price construction, and their analyses is based on duality theory for stochastic convex programs. Therefore, the results are invariant to the nature of the power flow relaxations adopted in the clearing models. More importantly, the duality theory seamlessly extends to multistage stochastic programs (see, e.g., \cite{higle2006multistage}), opening a viable path to design a hierarchical or a multi-settlement market that is consistent with operations planning prevalent in practice today. We will undertake a detailed study of multi-settlement markets in our future research. 

\begin{appendix}
\section{Detailed stochastic programming models} \label{sec:appendix_detailedForm}
This section presents the detailed problem formulations for the three stochastic programming (SP) formulations for the stochastic electric market clearing problem discussed in this paper. The notation used in these formulations is summarized in the following table.
\begin{table}[h!]
\begin{tabular}{|cl|}
\hline
\multicolumn{2}{|l|}{Sets} \\ 
\hline
\multicolumn{1}{|c|}{$\set{N}$}                     & Set of nodes (buses). \\ 
\hline
\multicolumn{1}{|c|}{$\set{L}$}                                             & Set of lines. \\ 
\hline
\multicolumn{1}{|c|}{$\set{P}$}                     & Set of market participants (generators an loads). \\ 
\hline
\multicolumn{1}{|c|}{$\Omega$}                                              & Set of scenarios. \\ 
\hline
\multicolumn{1}{|l|}{Parameters}                    & \\ 
\hline
\multicolumn{1}{|c|}{$c_i$}                         & Day-ahead biding price for $i^{th}$ participant. \\
\hline
\multicolumn{1}{|c|}{$\delta_i^+$/$\delta_i^-$}     & Real-time positive/negative deviation penalty of the $i^{th}$ participant.\\ 
\hline
\multicolumn{1}{|c|}{$x_i^{min}$/$x_i^{max}$}       & Day-ahead minimum/maximum capacity of the $i^{th}$ participant. \\
\hline
\multicolumn{1}{|c|}{$X_i^{min}$/$X_i^{max}$}       & Real-time minimum/maximum capacity of the $i^{th}$ participant \\ 
\hline
\multicolumn{1}{|c|}{$f_{ij}^{min}$/$f_{ij}^{max}$} & Day-ahead minimum/maximum capacity of the line $(i,j)$ \\ 
\hline
\multicolumn{1}{|c|}{$F_{ij}^{min}$/$F_{ij}^{max}$} & Real-time minimum/maximum capacity of the line $(i,j)$ \\ 
\hline
\end{tabular}
\end{table}

We denote the day-ahead settlement/cleared amounts by the decision variable $x_i$ for either type of participants $i \in \set{P}$. A decision variable corresponding to a generator satisfies $x_i \geq 0$, and that corresponding to a demand satisfies $x_i \leq 0$. In addition to the day-ahead clearing decisions, we denote the power flows that support the cleared day-ahead quantities by $f_{ij}$ for $(i,j) \in \set{L}$. The corresponding real-time decision variables are denoted $X_i$ and $F_{ij}$, respectively. In our models, we use $\set{F}$ to capture the feasible power flows in the network. While our analysis and pricing mechanisms apply to any convex approximation or relaxations of the power flows, we use the direct-current approximation in our experiments. In this case, the set is defined as 
\begin{align*}
\set{F} = \left \{(f_{ij})_{(i,j) \in \set{L}}, (\theta_i)_{i \in \set{B}}) \left \vert 
	\begin{array}{l}
		f_{ij} = \frac{V_iV_j}{B_{ij}}(\theta_i - \theta_j) \quad (i,j) \in \set{L}, \\ 
		\theta^{\min} \leq \theta_i \leq \theta^{\max}
	\end{array}\right. \right \}.
\end{align*}
Here, $\theta_i$ is an auxiliary decision variable that captures bus voltage angle at $i \in \set{B}$ . The parameter $V_i$ is the bus voltage magnitude which is usually set to one for all $i \in \set{B}$, and $B_{ij}$ is the line susceptance for $(i,j) \in \set{L}$. Finally, we use $\set{P}(n) \subset \set{P}$ to denote the subset of market participants that are connected to bus $n$.

\subsection{Canonical Formulation} \label{subsec:appendix_canonicalForm}
Using the above notation, the two-stage canonical formulation of the electricity market clearing is stated as follows:
\begin{subequations}\allowdisplaybreaks
\begin{align}
    \min ~ &\sum_{i \in \set{P}} c_ix_i + \expect{ (c_i+\delta^+_i) (X_i(\rv)-x_i)_+ - (c_i-\delta^-_i)(X_i(\rv) - x_i)_-}{}, \label{eq:canonical_obj}\\
    \text{s.t.}~ &x^{\min}_i \leq x_i \leq x^{\max}_i \qquad \forall i \in \set{P}, \label{eq:canonical_participant_cap}\\
    &f^{\min}_{ij} \leq f_{ij} \leq f^{\max}_{ij} \qquad \forall (i,j) \in \set{L}, \label{eq:canonical_line_cap}\\
    &(f_{ij})_{(i,j) \in \set{L}} \in \set{F}, \label{eq:canonical_line_phy}\\
    &\sum_{j:(j,n) \in \set{L}}f_{jn} - \sum_{j:(n,j) \in \set{L}}f_{jn} + \sum_{i \in \set{P}(n)}x_i = 0, \quad \forall n \in \set{N}, \label{eq:canonical_flowbalance}\\
    &X^{\min}_i \leq X_i(\rv) \leq \min\{X^{\max}_i, X^{\text{avail}}_i(\rv)\} \qquad \forall i \in \set{P}, \rv \in \Omega, \label{eq:canonical_rt_participant_cap}\\
    &F^{\min}_{ij} \leq F_{ij}(\rv) \leq F^{\max}_{ij} \qquad \forall (i,j) \in \set{L}, \rv \in \Omega, \label{eq:canonical_rt_line_cap}\\
    &(F_{ij}(\rv))_{(i,j) \in \set{L}} \in \set{F}, \qquad \forall \rv \in \Omega, \label{eq:canonical_rt_line_phy}\\
    &\sum_{j:(j,n) \in \set{L}}F_{jn}(\rv) - \sum_{j:(n,j) \in \set{L}}F_{jn}(\rv) - \sum_{j:(j,n) \in \set{L}}f_{jn} - \sum_{j:(n,j) \in \set{L}}f_{jn} \notag \\
    &\hspace{4cm}+ \sum_{i \in \set{P}(n)}(X_i(\rv) - x_i) = 0, \qquad \forall i \in \set{N}, \rv \in \Omega. \label{eq:canonical_rt_flowbalance}
\end{align}
\label{eq:canonical_model}
\end{subequations}
Notice that the model incorporates the stochastic behavior of the real-time market and uses an expectation-valued objective function, \eqref{eq:canonical_obj}. Constraints \eqref{eq:canonical_participant_cap}-\eqref{eq:canonical_flowbalance} represent the day-ahead market constraints. Constraints \eqref{eq:canonical_rt_participant_cap}-\eqref{eq:canonical_rt_flowbalance} are for the real-time market. Notice that each constraint is applied for every scenario $\rv \in \Omega$. When $\Omega$ has continuous support, the $\rv \in \Omega$ must be viewed as constraint being applied almost surely.

\subsection{Mean-vector form two-stage model} \label{subsec:appendix_meanvector}
The mean-vector formulation of the stochastic electricity clearing problem is stated as follows:
\begin{subequations} \allowdisplaybreaks
\begin{align} 
    \min ~ &\sum_{i \in \set{P}} \expect{c_ix_i(\rv) + (c_i+\delta^+_i) (X_i(\rv)-x_i(\rv))_+ - (c_i+\delta^-_i)(X_i(\rv) - x_i(\rv))_-}{}, \label{eq:meanVector_obj}\\
    \text{s.t.}~ &x^{\min}_i \leq x_i(\rv) \leq x^{\max}_i \qquad \forall i \in \set{P}, \label{eq:meanVector_participant_cap}\\
    &f^{\min}_{ij} \leq f_{ij}(\rv) \leq f^{\max}_{ij} \qquad \forall (i,j) \in \set{L}, \label{eq:meanVector_line_cap}\\
    &(f_{ij}(\rv))_{(i,j) \in \set{L}} \in \set{F}, \label{eq:meanVector_line_phy}\\
    &\sum_{j:(j,n) \in \set{L}}f_{jn} - \sum_{j:(n,j) \in \set{L}}f_{jn} + \sum_{i \in \set{P}(n)}x_i = 0, \quad \forall n \in \set{N}, \label{eq:meanVector_flowbalance}\\
    &X^{\min}_i \leq X_i(\rv) \leq \min\{X^{\max}_i, X^{\text{avail}}_i(\rv)\} \qquad \forall i \in \set{P}, \rv \in \Omega, \label{eq:meanVector_rt_participant_cap}\\
    &F^{\min}_{ij} \leq F_{ij}(\rv) \leq F^{\max}_{ij} \qquad \forall (i,j) \in \set{L}, \rv \in \Omega, \label{eq:meanVector_rt_line_cap}\\
    &(F_{ij}(\rv))_{(i,j) \in \set{L}} \in \set{F}, \qquad \forall \rv \in \Omega, \label{eq:meanVector_rt_line_phy}\\
    &\sum_{j:(j,n) \in \set{L}}F_{jn}(\rv) - \sum_{j:(n,j) \in \set{L}}F_{jn}(\rv) - \sum_{j:(j,n) \in \set{L}}f_{jn} - \sum_{j:(n,j) \in \set{L}}f_{jn} \notag \\
    &\hspace{4cm}+ \sum_{i \in \set{P}(n)}(X_i(\rv) - x_i) = 0, \qquad \forall i \in \set{N}, \rv \in \Omega, \label{eq:meanVector_rt_flowbalance}\\
    &x_i(\rv) - \expect{x_i(\rv)}{} = 0 \qquad \forall i \in \set{P}, \rv \in \Omega, \label{eq:meanVector_NA_participants}\\
    &f_{ij}(\rv) - \expect{f_{ij}(\rv)}{} = 0 \qquad \forall (i,j) \in \set{L}, \rv \in \Omega. \label{eq:meanVector_NA_flow}
\end{align}
\label{eq:meanVector_model}
\end{subequations}
The non-anticipativity constraints are given in \eqref{eq:meanVector_NA_participants} and \eqref{eq:meanVector_NA_flow} for clearing amounts and power flows respectively. In these equations, the scenario-dependent day-ahead solution is set to their respective expected values.

\subsection{State-vector form two-stage model} \label{subsec:appendix_statevector}
The state-vector two-stage SP model is stated as follows:
\begin{subequations} \allowdisplaybreaks
\begin{align} 
    \min ~ &\sum_{i \in \set{P}} \expect{c_ix_i(\rv) + (c_i+\delta^+_i) (X_i(\rv)-x_i(\rv))_+ - (c_i+\delta^-_i)(X_i(\rv) - x_i(\rv))_-}{}, \label{eq:stateVector_obj}\\
    \text{s.t.}~ &x^{\min}_i \leq x_i(\rv) \leq x^{\max}_i \qquad \forall i \in \set{P}, \label{eq:stateVector_participant_cap}\\
    &f^{\min}_{ij} \leq f_{ij}(\rv) \leq f^{\max}_{ij} \qquad \forall (i,j) \in \set{L}, \label{eq:stateVector_line_cap}\\
    &(f_{ij}(\rv))_{(i,j) \in \set{L}} \in \set{F}, \label{eq:stateVector_line_phy}\\
    &\sum_{j:(j,n) \in \set{L}}f_{jn} - \sum_{j:(n,j) \in \set{L}}f_{jn} + \sum_{i \in \set{P}(n)}x_i = 0, \quad \forall n \in \set{N}, \label{eq:stateVector_flowbalance}\\
    &X^{\min}_i \leq X_i(\rv) \leq \min\{X^{\max}_i, X^{\text{avail}}_i(\rv)\} \qquad \forall i \in \set{P}, \rv \in \Omega, \label{eq:stateVector_rt_participant_cap}\\
    &F^{\min}_{ij} \leq F_{ij}(\rv) \leq F^{\max}_{ij} \qquad \forall (i,j) \in \set{L}, \rv \in \Omega, \label{eq:stateVector_rt_line_cap}\\
    &(F_{ij}(\rv))_{(i,j) \in \set{L}} \in \set{F}, \qquad \forall \rv \in \Omega, \label{eq:stateVector_rt_line_phy}\\
    &\sum_{j:(j,n) \in \set{L}}F_{jn}(\rv) - \sum_{j:(n,j) \in \set{L}}F_{jn}(\rv) - \sum_{j:(j,n) \in \set{L}}f_{jn} - \sum_{j:(n,j) \in \set{L}}f_{jn} \notag \\
    &\hspace{4cm}+ \sum_{i \in \set{P}(n)}(X_i(\rv) - x_i) = 0,, \qquad \forall i \in \set{N}, \rv \in \Omega, \label{eq:stateVector_rt_flowbalance}\\
    &x_i(\rv) - \chi^x_i = 0 \qquad \forall i \in \set{P}, \rv \in \Omega, \label{eq:stateVector_NA_participants}\\
    &f_{ij}(\rv) - \chi^f_{ij} = 0 \qquad \forall (i,j) \in \set{L}, \rv \in \Omega. \label{eq:stateVector_NA_flow}
\end{align}
\label{eq:stateVector_model}
\end{subequations}
Notice that additional non-anticipative constraints \eqref{eq:stateVector_NA_participants} and \eqref{eq:stateVector_NA_flow} for day-ahead dispatch amounts and power flows, respectively. In these equations, the scenario-dependent quantities are set equal to their respective state variables.

\section{Omitted Proofs} \label{sec:proofs}
This section presents the proof of the technical results concerning the properties of pricing mechanisms \ref{eq:Pricing_mechanism_C}, \ref{eq:Pricing_mechanism_MV}, and \ref{eq:Pricing_mechanism_SV}. The proofs for Propositions \ref{prop:meanVectorDual} and \ref{prop:stateVectorDual} are based on the corresponding results for multistage stochastic convex programs in \cite{higle2006multistage}. Since we are focused on two-stage formulations in this paper, our results in these propositions are a special case of the more general settings studied in \cite{higle2006multistage}. 

\ifpaper \proof{Proof of Proposition \ref{prop:canonicalProperties}.} \else \begin{proof}[Proof of Proposition \ref{prop:canonicalProperties}.] \fi
    See Theorems 1 and 2 in \cite{pritchard2010single} for the results on revenue adequacy and cost recovery in expectation. Under pricing-mechanism \eqref{eq:Pricing_mechanism_C}, scenario-specific price distortion is given by $\mathbb{M}_n^c(\obs) = \pi_n^c - \Pi_n^c(\obs),~\forall n \in \set{N}, \obs \in \Omega$. Theorem 9 in \cite{zavala2017stochastic} establishes the bounds on expected price distortion, i.e.,$-\delta^+_i \leq \expect{\mathbb{M}^c_n(i)(\rv)}{} \leq \delta^-_i ,~\forall i \in \set{P}$. 
\ifpaper \hfill \Halmos \endproof \else \end{proof} \fi

The proofs of Proposition \ref{prop:meanVectorDual} and Proposition \ref{prop:stateVectorDual} follow similar agruments. Therefore, we only construct the necessary elements to prove Proposition \ref{prop:meanVectorDual} in the following. 
\ifpaper \proof{Proof of } \else \begin{proof}[Proof of Proposition \ref{prop:meanVectorDual}.] \fi
To begin, note that for any $\y_1 \in \set{L}_{n_1}^\infty$ and $\mu \in \set{L}_{n_1}^1$, we have $\expect{\inner{\mu(\rv), \expect{\y_1(\rv)}{\mathbb{P}}}}{\mathbb{P}} = \inner{\expect{\mu(\rv)}{\mathbb{P}}, \expect{\y_1(\rv)}{\mathbb{P}}} = \expect{\inner{\expect{\mu(\rv)}{\mathbb{P}}, \y_1(\rv)}}{\mathbb{P}}$. Therefore, 
\begin{align}
    \expect{\inner{\mu(\rv), \y_1(\rv) - \expect{\y_1(\rv)}{\mathbb{P}}}}{\mathbb{P}} \notag
    =~& \expect{\inner{\mu(\rv), \y_1(\rv)} - \inner{\mu(\rv), \expect{\y_1(\rv)}{\mathbb{P}}}}{\mathbb{P}} \notag \\
    =~& \expect{\inner{\mu(\rv), \y_1(\rv)} - \inner{\expect{\mu(\rv)}{\mathbb{P}}, \y_1(\rv)}}{\mathbb{P}} \notag \\
    =~& \expect{\inner{\mu(\rv) - \expect{\mu(\rv)}{\mathbb{P}}, \y_1(\rv)}}{\mathbb{P}}. \label{eq:LagrangianEquiv}
\end{align}
We define a stochastic Lagrangian function $\mathbb{L}(\mu, \obs) = \sup_{\y_1 \in \RR^{n_1}} \{\inner{\mu(\obs) - \expect{\mu(\rv)}{\mathbb{P}}, \y_1} - \phi(\y_1, \obs)\}$. Using the Lagrangian and the relationship in \eqref{eq:LagrangianEquiv}, we obtain a dual form of the mean-vector formulation which satisfies: 
\begin{align}
    v_p^\star \geq \sup_{\mu \in \set{L}_{n_1}^1} -\expect{L(\mu(\rv), \rv)}{\mathbb{P}} = \sup_{\mu \in \set{L}_{n_1}^1} \expect{\phi(\y_1(\rv), \rv) + \inner{\mu(\rv) - \expect{\mu(\rv)}{\mathbb{P}}, \y_1(\rv)}}{\mathbb{P}}. % \equiv \sup_{\mu \in \set{L}_{n_1}^1} -\expect{\phi^\star(\mu(\rv) - \expect{\mu(\rv)}{\mathbb{P}}}{\mathbb{P}}.
\end{align}
Notice that the dual on the right-hand side is an unconstrained optimization problem. Using the stochastic Lagrangian function and the definition of a conjugate function, we have $\mathbb{L}(\mu,\obs) = \phi^*(\mu(\obs) - \expect{\mu(\rv)}{}, \obs)$. The rest of the proof follows the same arguments as Proposition \ref{prop:stateVectorDual} presented later in this section. 
\ifpaper \hfill \Halmos \endproof \else \end{proof} \fi

\ifpaper \proof{Proof of Theorem \ref{thm:meanVector_properties}.} \else \begin{proof}[Proof of Theorem \ref{thm:meanVector_properties}.] \fi
    Consider the Lagrangian relaxation of \eqref{eq:meanVectorPrimal}. 
    \begin{align}
        \mathbb{L}^m =~ &\Expect{\sum\limits_{i \in \set{P}}c_ix_i(\rv) + (c_i + \delta^+_i)(X_i(\rv) - x_i(\rv))_+ - (c_i - \delta^-_i)(X_i(\rv) - x_i(\rv))_- \notag \\ 
        &- \sum\limits_{n \in \set{N}}\pi_n(\rv) \big(\tau_n(f(\rv)) + \sum\limits_{i \in P(n)}x_i(\rv) \big) \notag \\
        &- \sum\limits_{n \in \set{N}}\Pi_n(\rv) \big(\tau_n(F(\rv)) - \tau_n(f(\rv)) + \sum\limits_{i \in P(n)}(X_i(\rv) - x_i(\rv)) \big) \notag \\
        &- \sum\limits_{i \in \set{P}} \mu^x_i(\rv) \big( x_i(\rv) - \expect{x_i(\rv)}{} \big) - \sum\limits_{i \in \set{L}} \mu^f_i(\rv) \big( f_i(\rv) - \expect{f_i(\rv)}{} \big)}{}. \label{eq:meanVector_lagrangian}
    \end{align}
    Since $\mathbb{L}^m$ is minimized over $\set{C}_1$ and $\set{C}_2$, the optimal solution ($x^*(\rv), X^*(\rv), f^*(\rv), F^*(\rv), \pi^*(\rv), \Pi^*(\rv), \mu^*(\rv)$) also minimizes the Lagrangian relaxation with the optimal value $\mathbb{L}^{m^*}$. Notice that the solution obtained by setting $f(\obs) = 0$ and $F(\obs) = 0$ for all $\obs \in \rvset$ is a sub-optimal feasible solution. For such a solution we have $\expect{f(\rv)}{} = 0$ and 
    \begin{align*}
    \mathbb{L}^{m^*} \leq &~ \Expect{\sum\limits_{i \in \set{P}}c_ix^*_i(\rv) + (c_i+\delta^+_i)(X^*_i(\rv) - x^*_i(\rv))_+ - (c_i - \delta^-_i)(X^*_i(\rv) - x^*_i(\rv))_-\\
    &- \sum\limits_{n \in \set{N}}\pi^*_n \big( \sum\limits_{i \in P(n)}x^*_i(\rv) \big) 
    - \sum\limits_{n \in \set{N}}\Pi^*_n(\rv) \big( \sum\limits_{i \in P(n)}(X^*_i(\rv) - x^*_i(\rv)) \big) \\
    &- \sum\limits_{i \in \set{P}} \mu^{x^*}_i(\rv) \big( x^*_i(\rv) - \expect{x^*_i(\rv)}{} \big) }{}.
    \end{align*}
    Substituting $\mathbb{L}^{m^*}$ from \eqref{eq:meanVector_lagrangian} and rearranging the terms we obtain
    \begin{align*}
        -\Expect{\sum\limits_{n \in \set{N}}\pi^*_n(\rv) \tau_n(f^*(\rv)) + \Pi^*_n(\rv) \big(\tau_n(F^*(\rv)) - \tau_n(f^*(\rv)) \big) + \sum\limits_{i \in \set{L}}\mu^{f^*}_i(\rv) \big( f^*_i(\rv) - \expect{f^*_i(\rv)}{} \big)}{} \leq 0.
    \end{align*}
    Since the optimal solution satisfies $f_i^*(\obs) - \expect{f_i^*(\rv)}{} = 0$ for all $i \in \set{L}, \obs \in \Omega$, the last term in the above inequality equates to zero. The optimal solution also satisfies $x^*_i(\obs) - \expect{x^*_i(\rv)}{} = 0$. Using this, we obtain
    \begin{align*}
        -\Expect{\sum\limits_{n \in \set{N}}\pi^*_n \tau_n(f) + \Pi_n^*(\rv) \big(\tau_n(F^*(\rv)) - \tau_n(f^*(\rv)) \big) + \sum\limits_{i \in \set{P}}\mu^{x^*}_i(\rv) \big( x^*_i(\rv) - \expect{x^*_i(\rv)}{} \big)}{} \leq 0.
    \end{align*}
    Using the flow balance equations \eqref{eq:da_flowBalance_original} and \eqref{eq:rt_flowBalance}, we have
    \begin{align*}
        \Expect{\sum\limits_{i \in \set{P}} \Big(\pi^*_{n(i)}(\rv)x^*_i(\rv) + \Pi^*_{n(i)}(\rv)(X^*_i(\rv) - x^*_i(\rv)) + \mu^{x^*}_i(\rv) \big( x^*_i(\rv) - \expect{x^*_i(\rv)}{} \big) \Big)}{} \leq 0.\\
        \Expect{\sum\limits_{i \in \set{P}} \Big(\pi^*_{n(i)}(\rv)x^*_i(\rv) + \Pi^*_{n(i)}(\rv)(X^*_i(\rv) - x^*_i(\rv)) + \mu^{x^*}_i(\rv) x^*_i(\rv)}{}\Big) - \sum\limits_{i \in \set{P}}\expect{\mu^{x^*}_i(\rv)}{} \expect{x^*_i(\rv)}{}  \leq 0.
    \end{align*}
    If we have $\sum_{i \in \set{P}}\expect{\mu^{x^*}_i(\rv)}{} \expect{x^*_i(\rv)}{} \leq 0$, we obtain $\expect{\sum_{i \in \set{P}} \rho^m_i(\rv)}{} \leq 0$ which implies that pricing-mechanism \eqref{eq:Pricing_mechanism_SV} is revenue adequate in expectation.
    
    % Cost recovery for generators
    Since the problem is convex, the optimal dual values, ($\pi^*(\rv), \Pi^*(\rv)$) satisfy
    \begin{align*}
        \mathbb{L}^{m^*} = \min\limits_{x(\rv),X(\rv), f(\rv),F(\rv)}~& 
        \Expect{\sum\limits_{i \in \set{P}}c_ix_i(\rv) + (c_i + \delta^+_i)(X_i(\rv) - x_i(\rv))_+ - (c_i - \delta^-_i)(X_i(\rv) - x_i(\rv))_- \\
        &- \sum\limits_{n \in \set{N}}\pi^*_n(\rv) \big(\tau_n(f(\rv)) + \sum\limits_{i \in P(n)}x_i(\rv) \big) \\
        &- \sum\limits_{n \in \set{N}}\Pi^*_n(\rv) \big(\tau_n(F(\rv)) - \tau_n(f(\rv)) + \sum\limits_{i \in P(n)}(X_i(\rv) - x_i(\rv)) \big) \\
        &- \sum\limits_{i \in \set{P}} \mu^{x^*}_i(\rv) \big( x_i(\rv) - \expect{x_i(\rv)}{} \big) - \sum\limits_{i \in \set{L}} \mu^{f^*}_i(\rv) \big( f_i(\rv) - \expect{f_i(\rv)}{} \big)}{}.
    \end{align*}
    Notice that the optimization problem on the right-hand side of the above inequality decomposes into participant-specific optimization problems. Therefore,
    \begin{align*}
        \mathbb{L}^{m^*} = \sum_{i \in \set{P}} \Expect{\min\limits_{x_i(\rv),X_i(\rv)}  \mathbb{L}^1_i(x_i(\rv), X_i(\rv))}{} +  \min\limits_{f(\rv),F(\rv)}\Expect{ \mathbb{L}^2(f(\rv), F(\rv))}{}.
     \end{align*}
    Here, we define
    \begin{align*}
        \mathbb{L}^1_i(x_i(\obs), X_i(\obs)) =& ~ c_ix_i(\obs) + (c_i + \delta^+_i)(X_i(\obs) - x_i(\obs))_+ - (c_i - \delta^-_i)(X_i(\obs) - x_i(\obs))_- - \pi^*_{n(i)}(\obs) x_i(\obs) \\
        &- \Pi^*_{n(i)}(\obs) (X_i(\obs) - x_i(\obs)) -  \mu^{x^*}_i(\obs) \big( x_i(\obs) - \expect{x_i(\rv)}{} \big) \quad \forall i \in \set{P},\\
        \mathbb{L}^2_i(f_i(\obs), F_i(\obs)) =& ~- \sum\limits_{n \in \set{N}}\pi^*_n(\obs) \tau_n(f(\obs)) - \sum\limits_{n \in \set{N}}\Pi^*_n(\obs) \big(\tau_n(F(\obs)) - \tau_n(f(\obs)) \big) \\
        &- \sum\limits_{i \in \set{L}} \mu^{f^*}_i(\obs) \big( f_i(\obs) - \expect{f_i(\rv)}{} \big).
    \end{align*}
Further, we can minimize $\mathbb{L}^{m^*}$ by minimizing the above functions separately for every participant. Now consider the Lagrangian $\mathbb{L}_i^1(x_i(\rv), X_i(\rv))$ and its optimal solution $(x^{*}_i(\rv), X^{*}(\rv))$. Since $(x^{\min}_i, X^{\min}_i)$ is a sub-optimal feasible solution, we have for all $i \in \set{P}$
    \begin{align*}
        \expect{\mathbb{L}_i^1(x^*_i(\rv), X^*_i(\rv))}{} \leq &~\Expect{c_ix^{\min}_i + (c_i + \delta^+_i)(X^{\min}_i - x^{\min}_i)_+ - (c_i - \delta^-_i)(X^{\min}_i - x^{\min}_i)_- \\
        &- \pi^*_{n(i)}(\rv) x^{\min}_i - \Pi^*_{n(i)}(\rv) (X^{\min}_i - x^{\min}_i) -  \mu^{x^*}_i(\rv) \big( x^{\min}_i - \expect{x^{\min}_i}{} \big)}{}.
    \end{align*}
    Substituting $\mathbb{L}_i^1(x^*_i(\rv), X^*_i(\rv)$ and rearranging the terms, we obtain
    \begin{align*}
        &\Expect{c_ix^*_i(\obs) + (c_i + \delta^+_i)(X^*_i(\obs) - x^*_i(\obs))_+ - (c_i - \delta^-_i)(X^*_i(\obs) - x^*_i(\obs))_- \\
        &\hspace{1.5cm}- \Big( c_ix^{\min}_i + (c_i + \delta^+_i)(X^{\min}_i - x^{\min}_i)_+ - (c_i - \delta^-_i)(X^{\min}_i - x^{\min}_i)_- \Big)}{} \\
        \leq ~& \Expect{\big( \pi^*_{n(i)}(\obs) + \mu^{x^*}_i(\obs) \big) x^*_i(\obs) + \Pi^*_{n(i)} (\obs) \big(X_i^*(\obs) - x_i^*(\obs) \big)}{} - \expect{\mu^{x^*}_i(\rv)}{}\expect{x^*_i(\rv)}{}\\
        &\hspace{1.5cm} - \Expect{ \pi^*_{n(i)}(\obs) + \mu^{x^*}_i(\obs) \big) x^{\min}_i + \Pi^*_{n(i)}(\obs) \big(X^{\min}_i - x^{\min}_i)}1{} + \expect{\mu^{x^*}_i(\rv)}{}\expect{x^{\min}_i(\rv)}{}.
\end{align*}
If the startup costs are sufficiently covered through the uplifts paid by the commitment problem, we can focus on payment settled using the clearing problem studied here. In this case, we have
\begin{align*}
	\Expect{c_ix^*_i(\obs) + &(c_i + \delta^+_i)(X^*_i(\obs) - x^*_i(\obs))_+ - (c_i - \delta^-_i)(X^*_i(\obs) - x^*_i(\obs))_-}{}  \\ 
	\leq ~& \Expect{\big( \pi^*_{n(i)}(\obs) + \mu^{x^*}_i(\obs) \big) x^*_i(\obs) + \Pi^*_{n(i)} (\obs) \big(X_i^*(\obs) - x_i^*(\obs) \big)}{} - \expect{\mu^{x^*}_i(\rv)}{}\expect{x^*_i(\rv)}{}.
\end{align*}
Using \eqref{eq:socialSurplus_part} and \eqref{eq:Pricing_mechanism_MV} in the above inequality, we obtain as $ \expect{\rho_i^m(\rv) - \varphi_i(\rv)}{} \geq 0$ whenever $\expect{\mu^{x^*}_i(\rv)}{}\expect{x^*_i(\rv)}{} \geq 0$. Noting that $\expect{\mu_i^{x^*}(\rv)}{} \geq 0$ for all $i \in \set{G}$, the latter condition holds for all generators implying that they recover their costs in expectation. 
    
% Price distortion
We next show the price distortion properties. Notice that
\begin{align*}
    c_ix_i(\obs) + (c_i + \delta^+_i)(X_i(\obs) - x_i(\obs))_+ - &(c_i - \delta^-_i)(X_i(\obs) - x_i(\obs))_-\\
    = ~& c_iX_i(\obs) + (\delta^+_i + \delta^-_i)(X_i(\obs)-x_i(\obs))_+ - \delta^-_i(X_i(\obs)-x_i(\obs))
\end{align*}
Using the above relationship in $\mathbb{L}^m$, the partial Lagrangian with respect to day-ahead quantity $x_i(\rv)$ at the stationary point can be written as follows:
    \begin{align*}
    &0 ~\in~ \partial_{x_i(\obs)}\mathbb{L}^m = (\delta^+_i + \delta^-_i)\partial_{x_i(\obs)}(X_i(\obs) - x_i(\obs))_+ + \delta^-_i - \big(\pi_{n(i)}(\obs) - \Pi_{n(i)}(\obs) + (1 - p(\obs))\mu^x_i(\obs)\big) \\
    &\Rightarrow \cfrac{- \delta^-_i + (\pi_{n(i)}(\obs) - \Pi_{n(i)}(\obs) + (1 - p(\obs))\mu^x_i(\obs))}{\delta^+_i + \delta^-_i} ~~ \in ~~ \partial_{x_i(\obs)}(X_i(\obs) - x_i(\obs))_+.
    \end{align*}
    Since, 
    \begin{align*}
    \partial_{x_i(\obs)}(X_i(\obs) - x_i(\obs))_+ = 
    \begin{cases}
        -1, & \text{if } X_i(\obs) > x_i(\obs)\\
        0, & \text{if } X_i(\obs) < x_i(\obs)\\
        [-1, 0] & \text{if } X_i(\obs) = x_i(\obs)
    \end{cases}.
    \end{align*}
    We have
    \begin{align*}
    & -1 \leq \cfrac{- \delta^-_i + (\pi_{n(i)}(\obs) - \Pi_{n(i)}(\obs) + (1 - p(\obs))\mu^x_i(\obs))}{\delta^+_i + \delta^-_i} \leq 0 \\
    & \Rightarrow -\delta^+_i + p(\obs)\mu^x_i(\obs) ~\leq~ (\pi_{n(i)}(\obs) + \mu^x_i(\obs)) - \Pi_{n(i)}(\obs)  ~\leq~ \delta^-_i + p(\obs)\mu^x_i(\obs) \qquad \forall i \in \set{P}, \obs \in \Omega.
\end{align*}
This completes the proof. \ifpaper \hfill \Halmos \endproof \else
\end{proof} \fi

To establish an appropriate dual problem for \eqref{eq:stateVectorPrimal}, we begin with the following result that establishes that $\set{M}$ is complementary to $\set{N}_\infty$.
\begin{lemma}\label{lemma:complementary}
	For any $\y_1 \in \set{N}_\infty$ and $\sigma \in \set{M}$, we have $\expect{\inner{\sigma(\rv), \y_1(\rv)}}{\mathbb{P}} = 0$.
\end{lemma}
\begin{proof}
    Since $\y_1(\rv) = \chi_1, a.s.$, we have $	\expect{\inner{\sigma(\rv), \y_1(\rv)}}{\mathbb{P}} =  \expect{\inner{\sigma(\rv), \chi_1}}{\mathbb{P}} =  \inner{\expect{\sigma(\rv)}{\mathbb{P}}, \chi_1} = 0$. 
\end{proof}

\ifpaper \proof{Proof of Proposition \ref{prop:stateVectorDual}.} \else \begin{proof}[Proof of Proposition \ref{prop:stateVectorDual}.] \fi
    Consider a feasible solution $(\bar{\y}_1, \bar{\chi}_1) \in \widehat{\set{N}}_\infty$ to \eqref{eq:stateVectorPrimal}. Using the conjugate of $\phi(\bar{\y}_1, \obs)$ we have 
    \begin{align*}\renewcommand{\arraystretch}{1}
	\begin{array}{rrl}
	& \phi^\star(\sigma(\obs), \obs) =& \sup_{\y_1 \in \RR^{n_1}}~ \big\{\inner{\sigma(\obs), \y_1(\obs)} - \phi(\y_1(\obs),\obs)\big\} \\
	& \geq & \inner{\sigma(\obs), \bar{\y}_1(\obs)} - \phi(\bar{\y}_1(\obs),\obs) \qquad \forall \obs \in \rvset. \\
	\Rightarrow & \quad \phi(\bar{\y}_1(\obs),\obs) \geq & -\phi^\star(\sigma(\obs),\obs) + \inner{\sigma(\obs), \bar{\y}_1(\obs)} \qquad \forall \obs \in \rvset. \\
	\Rightarrow & \quad \expect{\phi(\bar{\y}_1(\rv),\rv)}{\mathbb{P}} \geq & -\expect{\phi^\star(\sigma(\rv), \rv)}{\mathbb{P}} + \expect{\inner{\sigma(\rv), \bar{\y}_1(\rv)}}{\mathbb{P}} \\
	\end{array}	
    \end{align*}
    The term on the right-hand side of the last inequality can attain a value of $+\infty$ unless $\expect{\sigma(\rv)}{\mathbb{P}} = 0$, in which case we have $\expect{\inner{\sigma(\rv), \bar{\y}_1(\rv)}}{\mathbb{P}} = 0$ from Lemma \ref{lemma:complementary}. Therefore, we can obtain a lower bound on the optimal value of \eqref{eq:stateVectorPrimal} as
    \begin{align} \label{eq:weakDuality}
    	v_p^\star \geq \sup_{\sigma \in \set{L}^1_{n_1}} \big\{-\expect{\phi^\star(\sigma(\rv), \rv)}{\mathbb{P}} ~|~ \expect{\sigma(\rv)}{\mathbb{P}} = 0\big\}.
    \end{align}

    Since the mapping $\obs \rightarrow \partial\phi(\bar{\y}_1(\obs),\obs)$ is $\mathbb{P}$-measurable and closed, there exists a $\bar{\sigma} \in \set{L}^1_{n_1}$ such that $\bar{\sigma}(\rv) \in \partial\phi(\bar{\y}_1(\rv),\rv)$, almost surely. This implies that
    \begin{align*} \renewcommand{\arraystretch}{1}
    	\begin{array}{rrl} 
    	&\phi(\y_1,\obs) \geq & \phi(\bar{\y}_1(\obs), \obs) + \inner{\bar{\sigma}(\obs), \y_1 - \bar{\y}_1(\obs)}  \qquad \forall \y_1 \in \RR^{n_1}, \obs \in \rvset.\\
    	\Rightarrow & \inner{\bar{\sigma}(\obs), \bar{\y}_1(\obs)} - \phi(\bar{\y}_1(\obs), \obs) \geq & \inner{\bar{\sigma}(\obs), \y_1} - \phi(\y_1,\obs) \qquad \forall \y_1 \in \RR^{n_1}, \obs \in \rvset.
    	\end{array}
    \end{align*}
    Therefore, $\bar{\y}_1(\rv) \in \argmax\{\inner{\bar{\sigma}(\rv), \y_1} - \phi(\y_1,\rv)\}$, almost surely. It follows from the definition of conjugate functions that the term on the left-hand side of the above inequality is $\phi^\star(\bar{\sigma}(\obs),\obs)$. This implies 
    \begin{align} \label{eq:squeeze}
    	- \expect{\phi^\star(\bar{\sigma}(\rv),\rv)}{\mathbb{P}} = \expect{\phi(\bar{\y}_1(\rv),\rv)}{\mathbb{P}} - \expect{\inner{\bar{\sigma}(\rv), \bar{\y}_1(\rv)}}{\mathbb{P}}.
    \end{align}
    Notice that \eqref{eq:squeeze} holds for any $\bar{\y}_1 \in \set{N}_\infty$ and the subgradient $\bar{\sigma}$ calculated at $\bar{\y}_1$.
    
    Now, if we focus specifically on $(\bar{\y}_1, \bar{\chi}_1)$ that is an optimal solution to \eqref{eq:stateVectorPrimal}, we have $ 0 \in \partial\expect{\phi(\bar{\y}_1(\rv),\rv)}{\mathbb{P}} + \widehat{\set{N}}_\infty^\bot(\bar{\y}_1, \bar{\chi}_1)$. Here, $\widehat{\set{N}}^\bot_\infty(\y_1, \chi_1)$ is the normal cone to $\widehat{\set{N}}_\infty$ at the point $(\y_1, \chi_1)$ such that any $(\eta_{\y}, \eta_\chi) \in \widehat{\set{N}}^\bot_\infty(\y_1, \chi_1)$ satisfies $\inner{\eta_\y(\rv), \y_1(\rv)} + \inner{\eta_\chi(\rv), \chi_1} = 0$, almost surely. This implies that%
    \begin{align}
    	0 =&~ \expect{\inner{\eta_\y(\rv), \y_1(\rv)} + \inner{\eta_\chi(\rv), \chi_1}}{\mathbb{P}} = \expect{\inner{\eta_\y(\rv) + \eta_\chi(\rv), \chi_1}}{\mathbb{P}} \notag \\
    	  =&~ \inner{\expect{\eta_\y(\rv) + \eta_\chi(\rv)}{\mathbb{P}}, \chi_1}. \notag \\
    	  \text{If}~ \chi_1 \neq 0 \Rightarrow \qquad &\expect{\eta_\y(\rv) + \eta_\chi(\rv)}{\mathbb{P}} = 0. \label{eq:na_normalCone1}
    \end{align}
    Further, since $\phi$ is a lower semicontinuous proper convex function on $\RR^{n_1}$, almost surely, we have $\partial\expect{\phi(\bar{\y}_1(\rv),\rv)}{\mathbb{P}} = \expect{\partial\phi(\bar{\y}_1(\rv),\rv)}{\mathbb{P}}$ \citep{rockafellar1982interchange}. The first-order optimality condition, therefore, implies that there exists $(\eta_\y, \eta_\chi) \in \widehat{\set{N}}^\bot_\infty(\bar{\y}_1, \bar{\chi}_1)$ that along with $\bar{\sigma}(\rv) \in \partial\phi(\bar{\y}_1(\rv),\rv)$ satisfies
    \begin{align}\label{eq:na_normalCone2}
    	\eta_\y(\rv) + \bar{\sigma}(\rv) = 0 ~\text{and}~ \eta_\chi(\rv) = 0\quad a.s. \quad\Rightarrow \quad \expect{\eta_\y(\rv) + \bar{\sigma}(\rv)}{\mathbb{P}} = 0 ~\text{and}~\expect{\eta_\chi(\rv)}{\mathbb{P}} = 0.
    \end{align}
    Combining the above result with \eqref{eq:na_normalCone1}, we conclude that $\expect{\bar{\sigma}(\rv)}{\mathbb{P}} = 0$. Using this result in \eqref{eq:squeeze} and Lemma \ref{lemma:complementary}, we obtain an upper bound on the optimal value of \eqref{eq:stateVectorPrimal}.
    \begin{align}
    	&- \expect{\phi^\star(\bar{\sigma}(\rv),\rv)}{\mathbb{P}} = \expect{\phi(\bar{\y}_1(\rv),\rv)}{\mathbb{P}} - \expect{\inner{\bar{\sigma}(\rv), \bar{\y}_1(\rv)}}{\mathbb{P}} = \expect{\phi(\bar{\y}_1(\rv),\rv)}{\mathbb{P}} = \min \eqref{eq:stateVectorPrimal} \notag \\
    	\Rightarrow &\sup_{\sigma \in \set{L}^1_{n_1}} \big\{-\expect{\phi^\star(\sigma(\rv), \rv)}{\mathbb{P}}~|~\expect{\sigma(\rv)}{\mathbb{P}} = 0 \big\} \geq - \expect{\phi^\star(\bar{\sigma}(\rv),\rv)}{\mathbb{P}} = \min \eqref{eq:stateVectorPrimal}. \label{eq:reverseDualInequality}
    \end{align}
    Combining \eqref{eq:weakDuality} and \eqref{eq:reverseDualInequality} completes the proof. 
\ifpaper \hfill \Halmos \endproof \else \end{proof} \fi

\ifpaper \proof{Proof of Proposition \ref{prop:infoPrice}.} \else \begin{proof}[Proof of Proposition \ref{prop:infoPrice}.] \fi
    Let $\bar{\sigma} \in \set{M}$ and $(\bar{\y}_1, \bar{\chi}_1) \in \widehat{\set{N}}_\infty$ be the minimizer to the Lagrangian $\mathbb{L}$. From the first-order optimality conditions, we have 
    \begin{align*} 
        0 \in \partial\mathbb{L}(\bar{\y}_1, \bar{\chi}_1, \bar{\sigma}) =~& (\expect{\partial \phi(\y_1(\rv), \rv)}{\mathbb{P}} + \expect{\bar{\sigma}(\rv)}{\mathbb{P}}, -\expect{\bar{\sigma}(\rv)}{\mathbb{P}})^\top.
    \end{align*}
    Since the set $\set{M}$ is equivalent to $\set{N}_\infty^\bot$, it follows that $0 \in \expect{\partial \phi(\y_1(\rv), \rv)}{\mathbb{P}} + \widehat{\set{N}}_\infty^\bot(\bar{\y}_1, \bar{\chi}_1)$, which is the sufficient condition for $(\bar{\y}_1, \bar{\chi}_1)$ to be an optimal solution for \eqref{eq:stateVectorPrimal}. The arguments for the converse follow from \eqref{eq:na_normalCone1} and \eqref{eq:na_normalCone2}. 
\ifpaper \hfill \Halmos \endproof \else \end{proof} \fi

\ifpaper \proof{Proof of Theorem \ref{thm:stateVector_properties}.} \else \begin{proof}[Proof of Theorem \ref{thm:stateVector_properties}.] \fi
    % Revenue adequacy in expectation
    Consider the following Lagrangian relaxation of the \eqref{eq:stateVectorPrimal}. 
    \begin{align}
        \mathbb{L}^s = &\Expect{\sum\limits_{i \in \set{P}}c_ix_i(\rv) + (c_i+\delta^+_i)(X_i(\rv) - x_i(\rv))_+ - (c_i-\delta^-_i)(X_i(\rv) - x_i(\rv))_- \notag \\
        &- \sum\limits_{n \in \set{N}}\pi_n(\rv) \big(\tau_n(f(\rv)) + \sum\limits_{i \in P(n)}x_i(\rv) \big) \notag \\
        &- \sum\limits_{n \in \set{N}}\Pi_n(\rv) \big(\tau_n(F(\rv)) - \tau_n(f(\rv)) + \sum\limits_{i \in P(n)}(X_i(\rv) - x_i(\rv)) \big) \notag \\
        &- \sum\limits_{i \in \set{P}} \sigma^x_i(\rv) \big( x_i(\rv) - \chi^x_i \big) - \sum\limits_{i \in \set{L}} \sigma^f_i(\rv) \big( f_i(\rv) - \chi_i^f \big)}{}. \label{eq:stateVector_lagrangian}
    \end{align}
    Since $\mathbb{L}^s$ is minimized over $\set{C}_1$ and $\set{C}_2$, the optimal solution ($x^*_i(\rv), X^*_i(\rv), \chi_i^{x^*}, f^*_i(\rv), F^*_i(\rv), \chi^{f^*}_i, \pi_i^*(\rv), \Pi_i^*(\rv), \sigma_i^{x^*}, \sigma_i^{f^*}$) also minimizes the Lagrangian relaxation. Let $\mathbb{L}^{s^*}$ be the value of the Lagrangian at the optimal solution. Further since, $(x^*_i(\rv), X^*_i(\rv), \chi_i^{x^*}, 0, 0, \chi_i^{f^*}, \pi_i^*(\rv), \Pi_i^*(\rv), \sigma_i^{x^*}, \sigma_i^{f^*})$ (obtained by setting $f(\obs) = 0$ and $F(\obs) = 0$ for all $\obs \in \rvset$) is a sub-optimal feasible solution, we have
    \begin{align*}
        \mathbb{L}^{s^*} \leq~& \Expect{\sum\limits_{i \in \set{P}} c_ix^*_i(\rv) + (c_i+\delta^+_i)(X^*_i(\rv) - x^*_i(\rv))_+ - (c_i-\delta^-_i)(X^*_i(\rv) - x^*_i(\rv))_- \\ - &\sum\limits_{n \in \set{N}}\pi^*_n(\rv) \big( \sum\limits_{i \in P(n)}x^*_i(\rv) \big) - \sum\limits_{n \in \set{N}}\Pi_n^*(\rv) \big( \sum\limits_{i \in P(n)}(X^*_i(\rv) - x^*_i(\rv)) \big) \notag \\
        &- \sum\limits_{i \in \set{P}} \sigma^{x^*}_i(\rv) \big( x^*_i(\rv) - \chi^{x^*}_i \big) + \sum\limits_{i \in \set{L}} \sigma^{f^*}_i(\rv) \chi^{f^*}_i}{}.
    \end{align*}
    Substituting $\mathbb{L}^{s^*}$ from \eqref{eq:stateVector_lagrangian} on the left-hand side of the above and rearranging the terms, we obtain
    \begin{align*}
        &-\Expect{\sum\limits_{n \in \set{N}}\pi^*_n(\rv) \tau_n(f^*(\rv)) + \Pi_n^*(\rv) \big(\tau_n(F^*(\rv)) - \tau_n(f^*(\rv)) \big) + \sum\limits_{i \in \set{L}}\sigma^{f^*}_i(\rv) f^*_i(\rv) }{} \leq 0.
    \end{align*}
    Since, $f_i^*(\obs) = \chi_i^{f^*} ~ \forall i \in \set{L}, \obs \in \Omega$ 
    \begin{align*}
        &-\expect{\sum\limits_{n \in \set{N}}\pi^*_n(\rv) \tau_n(f^*(\rv)) + \Pi^*_n(\rv) \big(\tau_n(F^*(\rv)) - \tau_n(f^*(\rv)) \big)}{} + \expect{ \sum\limits_{i \in \set{L}}\sigma^{f^*}_i(\rv)}{} \chi^{f^*}_i \leq 0,\\
       \Rightarrow  &-\expect{\sum\limits_{n \in \set{N}}\pi^*_n(\rv) \tau_n(f^*(\rv)) + \Pi^*_n(\rv) \big(\tau_n(F^*(\rv)) - \tau_n(f^*(\rv)) \big)}{} \leq 0.
    \end{align*}
    In the second inequality, we have used $\expect{ \sum\limits_{i \in \set{L}}\sigma^{f^*}_i(\rv)}{} = 0$, due to Theorem \ref{thm:solutionRelationship} $\ref{thm:solutionRelationship_NA_duals}$ Similarly, since $\Expect{ \sum_{i \in \set{P}}\sigma^{x^*}_i(\rv)}{} = 0$ (also from Theorem \ref{thm:solutionRelationship} $\ref{thm:solutionRelationship_NA_duals}$), we have
    \begin{align*}
        &-\Expect{\sum\limits_{n \in \set{N}}\pi^*_n(\rv) \tau_n(f^*(\rv)) + \Pi^*_n(\rv) \big(\tau_n(F^*(\rv)) - \tau_n(f^*(\rv)) \big)}{} + \expect{ \sum\limits_{i \in \set{P}}\sigma^{x^*}_i(\rv)\chi^{x^*}_i}{} \leq 0.
    \end{align*}
    Using the flow balance equations \eqref{eq:da_flowBalance_original} and \eqref{eq:rt_flowBalance}, and the state-vector form of the non-anticipativity constraint \eqref{eq:stateVector_na} in the above inequality, we obtain
    \begin{align*}
        &\Expect{\sum\limits_{i \in \set{P}} \big (\pi^*_{n(i)}(\rv) + \sigma^{x^*}_i(\rv))x_i(\rv) + \Pi^*_{n(i)}(\rv)(X_i^*(\rv) - x_i^*(\rv)) \big)}{} \leq 0.
    \end{align*}
    The above implies that $\expect{\sum_{i \in \set{P}} \rho^s_i(\rv)}{} \leq 0$, thereby, establishing the revenue adequacy of \eqref{eq:Pricing_mechanism_SV}.
    
    % Cost recovery for generators
    Since the problem is convex, the optimal dual values, ($\pi^*(\rv), \Pi^*(\rv), \sigma_i^{x^*}, \sigma_i^{f^*}$) satisfy
    \begin{align*}
        \mathbb{L}^{s^*} = \min\limits_{\substack{x(\rv),X(\rv), s^x,\\ f(\rv),F(\rv), s^f}}~& 
        \Expect{\sum\limits_{i \in \set{P}}c_ix_i(\rv) + (c_i+\delta^+_i)(X_i(\rv) - x_i(\rv))_+ - (c_i-\delta^-_i)(X_i(\rv) - x_i(\rv))_- \\
        &- \sum\limits_{n \in \set{N}}\pi^*_n(\rv) \big(\tau_n(f(\rv)) + \sum\limits_{i \in P(n)}x_i(\rv) \big) \\
        &- \sum\limits_{n \in \set{N}}\Pi^*_n(\rv) \big(\tau_n(F(\rv)) - \tau_n(f(\rv)) + \sum\limits_{i \in P(n)}(X_i(\rv) - x_i(\rv)) \big) \\
        &- \sum\limits_{i \in \set{P}} \sigma^{x^*}_i(\rv) \big( x_i(\rv) - \chi^{x}_i \big) - \sum\limits_{i \in \set{L}} \sigma^{f^*}_i(\rv) \big( f_i(\rv) - \chi^{f}_i \big)}{}.
    \end{align*}
    Notice that the optimization problem on the right-hand side of the above inequality decomposes into participant-specific optimization problems. Therefore,
    \begin{align*}
        \mathbb{L}^{s^*} = \sum_{i \in \set{P}} \Expect{\min\limits_{x_i(\rv),X_i(\rv), \chi_i^x}  \mathbb{L}^1_i(x_i(\rv), X_i(\rv), \chi^{x}_i)}{} +  \min\limits_{f(\rv),F(\rv), \chi^f}\expect{ \mathbb{L}^2(f(\rv), F(\rv), \chi^f)}{},
    \end{align*}
    where, we define
    \begin{align*}
        \mathbb{L}^1_i(x_i(\obs), &X_i(\obs), \chi^{x}_i) =~ c_ix_i(\obs) + (c_i+\delta^+_i)(X_i(\obs) - x_i(\obs))_+ - (c_i-\delta^-_i)(X_i(\obs) - x_i(\obs))_- \\
        &- \pi^*_{n(i)}(\obs) x_i(\obs) - \Pi^*_{n(i)}(\obs) (X_i(\obs) - x_i(\obs)) -  \sigma^{x^*}_i(\obs) \big( x_i(\obs) - \chi^{x}_i \big) \quad \forall i \in \set{P},\\
        \mathbb{L}^2(f_i(\obs), &F_i(\obs), \chi_i^f) =~- \sum\limits_{n \in \set{N}}\pi^*_n(\obs) \tau_n(f(\obs)) - \sum\limits_{n \in \set{N}}\Pi^*_n(\obs) \big(\tau_n(F(\obs)) - \tau_n(f(\obs)) \big) \\
        &- \sum\limits_{i \in \set{L}} \sigma^{f^*}_i(\obs) \big( f_i(\obs) - \chi^{f}_i \big),
    \end{align*}
    for all $\obs \in \rvset$. Further, we can minimize $\mathbb{L}^{s^*}$ by minimizing the above functions separately for every participant and $\obs \in \rvset$. Now consider the Lagrangian $\mathbb{L}^1(x_i(\obs), X_i(\obs), \chi^{x}_i)$ and its optimal solution $(x^{*}_i(\obs), X^{*}(\obs), \chi^{x^*}_i)$. Since $(x^{\min}_i, X^{\min}_i, \chi_i^{x^*})$ is a sub-optimal feasible solution, we have
    \begin{align*}
        \mathbb{L}(x^*_i(\obs), X^*_i(\obs), \chi^{x^*}_i) \leq &~c_ix^{\min}_i + (c_i+\delta^+_i)(X^{\min}_i - x^{\min}_i)_+ - (c_i-\delta^-_i)(X^{\min}_i - x^{\min}_i)_- \\
        &- \pi^*_{n(i)}(\obs) x^{\min}_i - \Pi^*_{n(i)}(\obs) (X^{\min}_i - x^{\min}_i) -  \sigma^{x^*}_i(\obs) \big( x^{\min}_i - \chi^{x^*}_i \big).
    \end{align*}
    Substituting $\mathbb{L}(x^*_i(\obs), X^*_i(\obs), \chi^{x^*}_i)$ and rearranging the terms, we obtain
    \begin{align*}
        &c_ix^*_i(\obs) + (c_i+\delta^+_i)(X^*_i(\obs) - x^*_i(\obs))_+ - (c_i-\delta^-_i)(X^*_i(\obs) - x^*_i(\obs))_- \\
        &\hspace{1.5cm}- \Big( c_ix^{\min}_i + (c_i+\delta^+_i)(X^{\min}_i - x^{\min}_i)_+ - (c_i-\delta^-_i)(X^{\min}_i - x^{\min}_i)_- \Big) \\
        \leq ~& \big( \pi^*_{n(i)}(\obs) + \sigma^{x^*}_i(\obs) \big) x_i(\obs) + \Pi^*_{n(i)} \big(\obs) (X_i(\obs) - x_i(\obs) \big)  \\
        &\hspace{1.5cm} - \Big( \pi^*_{n(i)}(\obs) + \sigma^{x^*}_i(\obs) \big) x^{\min}_i + \Pi^*_{n(i)}(\obs) \big(X^{\min}_i - x^{\min}_i \Big).
    \end{align*}
    Once again, if the uplifts payments from the unit commitment problem cover the minimum generation then the above can reduce to $ \varphi_i(\obs) \leq \rho_i^{s}(\obs) $. This implies cost recovery for all generators under every scenario $\obs \in \rvset$.

    % Price distortion
    Finally, to establish the property of scenario-specific price distortion, notice that
    \begin{align*}
        c_ix_i(\obs) + (c_i + \delta^+_i)(X_i(\obs) - x_i(\obs))_+ - &(c_i - \delta^-_i)(X_i(\obs) - x_i(\obs))_-\\
        = ~& c_iX_i(\obs) + (\delta^+_i + \delta^-_i)(X_i(\obs)-x_i(\obs))_+ - \delta^-_i(X_i(\obs)-x_i(\obs))
    \end{align*}
    Using the above relation in the Lagrangian $\mathbb{L}^s$ in \eqref{eq:stateVector_lagrangian}, the partial derivative of the Lagrangian with respect to day-ahead quantity $x_i(\obs)$ at the stationary point can be written as follows:
    \begin{align*}
        (\delta^+_i + \delta^-_i)\partial_{x_i(\obs)}(X_i(\obs) - x_i(\obs))_+ + \delta^-_i - (\pi_{n(i)}(\obs) - \Pi_{n(i)}(\obs) + \sigma^x_i(\obs)).
    \end{align*}
    Since $0 \in~ \partial_{x_i(\obs)}\mathbb{L}^s$, we have
    \begin{align*}
        &\Rightarrow \cfrac{- \delta^-_i + (\pi_{n(i)}(\obs) - \Pi_{n(i)}(\obs) + \sigma^x_i(\obs))}{\delta^+_i + \delta^-_i} ~~ \in ~~ \partial_{x_i(\obs)}(X_i(\obs) - x_i(\obs))_+.
    \end{align*}
    Note that
    \begin{align*}
        &\partial_{x_i(\obs)}(X_i(\obs) - x_i(\obs))_+ = 
        \begin{cases}
            -1, & \text{if } X_i(\obs) > x_i(\obs)\\
            0, & \text{if } X_i(\obs) < x_i(\obs)\\
            [-1, 0] & \text{if } X_i(\obs) = x_i(\obs).
        \end{cases}
    \end{align*}
    This implies that
    \begin{align*}
        & -1 \leq \cfrac{- \delta^-_i + (\pi_{n(i)}(\obs) - \Pi_{n(i)}(\obs) + \sigma^x_i(\obs))}{\delta^+_i + \delta^-_i} \leq 0, \\
        & \Rightarrow -\delta^+_i ~\leq~ \pi_{n(i)}(\obs) + \sigma^x_i(\obs) - \Pi_{n(i)}(\obs)  ~\leq~ \delta^-_i \qquad \forall i \in \set{P}, \obs \in \Omega.
    \end{align*}
    This completes the proof. 
\ifpaper \hfill \Halmos \endproof \else \end{proof} \fi

\section{Details of Test Systems used in computational study}\label{sec:testSystems} 
This section summarizes the characteristics of the three test systems used in our numerical experiments. 

\subsection{PZP-6}
\begin{figure}[t!]
\centering
    \includegraphics[scale = 0.6]{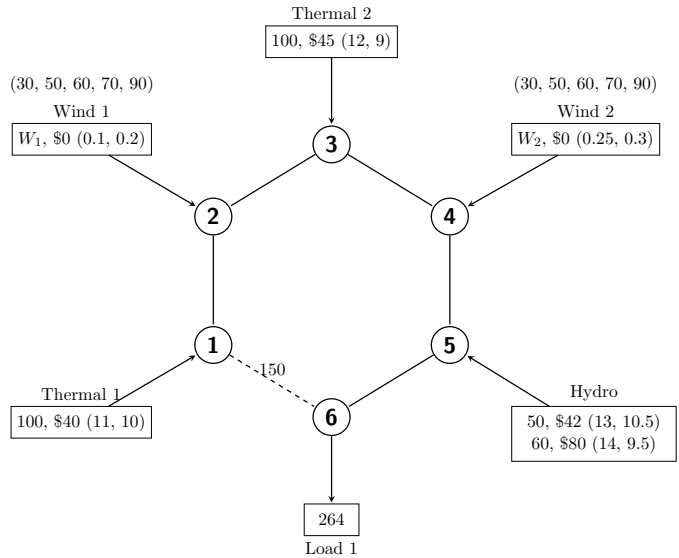}
    \caption{Network diagram of PZP-6 system.}
    \label{fig:6_node}
\end{figure}

We adopt the PZP-6 system from \cite{pritchard2010single}, depicted in Figure \ref{fig:6_node}. The system operates six buses connecting six generators and a load. The line connecting buses 1 and 6 has a maximum capacity of $150$ MW, and the remaining lines are uncapacitated. Figure \ref{fig:6_node} presents the capacities, day-ahead unit bid prices, real-time negative deviation and positive deviation premiums (in parentheses) for generators, and demand quantity for the load.
With equal probability, each wind generator can realize any scenario from the possible set $(30, 50, 60, 70, 90)$ (all in MW). Thus, there are a total of $25$ scenarios. Other generators can increase or decrease their generation in the real-time market to mitigate the impact of the wind generation uncertainty. Both hydro generators connect to the system through the same bus six. Further, we assume all lines are lossless. Note that we used 90, the maximum possible output, as the first-stage capacity limit of the wind generators.

\subsection{ZKAB-6}
\begin{figure}[t!]
\centering
    \includegraphics[scale = 0.6]{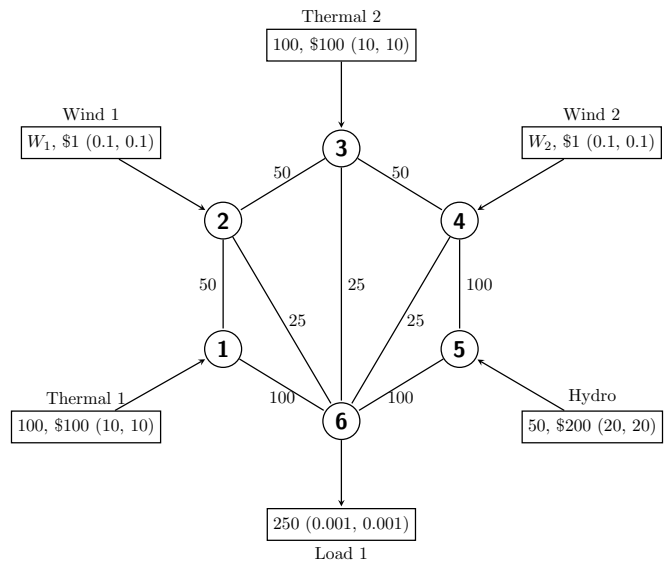}
    \caption{Network diagram of ZKAB-6 system.}
    \label{fig:zavala_6}
\end{figure}

The ZKAB-6 is a modified version of PZP-6 from \cite{zavala2017stochastic}, depicted in Figure \ref{fig:zavala_6}. Similar to PZP-6, the ZKAB-6 system operates six buses. However, the system has only five generators and a single deterministic load. All lines in the network have capacities, and they are shown in Figure \ref{fig:zavala_6}. Further, generator capacities, day-ahead unit bid prices, real-time negative deviation premiums and positive deviation premiums (in parentheses), and demand quantity for the load are shown in Figure \ref{fig:zavala_6}.
As in the case of PZP-6, each wind generator can realize any scenario from the possible set $(10, 20, 60, 70, 90)$ (all in MW), resulting in a total of $25$ scenarios. We assume all lines are lossless. Note that we used 90, the maximum possible output, as the first-stage capacity limit of the wind generators.

\subsection{SODA-30}
Our final test system is the IEEE 30 bus system. This network comprises 30 buses, six generators (4 thermal and two wind), and 21 demands. $\$200, \$175, \$100$, and $\$300$ are the generation costs for thermal generators, and renewable generation cost is $\$1$. Similarly, the positive and negative deviation premiums are $\$0.4, \$0.35, \$0.2, \$0.6$ for the thermal generators, and $\$0.65, \$0.6$ for the wind generators. We developed 200 independent scenarios for the wind electricity outputs. Further, all demands are deterministic. Similar to the previous instances, we assumed all the power lines are lossless, and the thermal generators can adjust their outputs in the real-time market. We set the day-ahead availabilities for the wind generators to their respective capacities. 

\end{appendix}

\bibliographystyle{apalike}
\bibliography{../../../StochasticClearing.bib}

\end{document}